\documentclass{emulateapj}
\usepackage{apjfonts}[9pt]

\begin{document}

\title{Massive Lyman Break Galaxies at $z\sim3$ in the 
\textit{Spitzer} Extragalactic First Look Survey}

\author{Hyunjin Shim\altaffilmark{1}, Myungshin Im\altaffilmark{1},
Phillip Choi\altaffilmark{2}, Lin Yan\altaffilmark{3},
 \& Lisa Storrie-Lombardi\altaffilmark{3}}
\altaffiltext{1}{Department of Physics \& Astronomy, FPRD, 
 Seoul National University, Seoul, Korea ; 
 hjshim@astro.snu.ac.kr, mim@astro.snu.ac.kr}
\altaffiltext{2}{Department of Physics \& Astronomy, Pomona College, 
 CA91711}
\altaffiltext{3}{Spitzer Science Center, California Institute of 
 Technology, CA91125}

\begin{abstract}
  We investigate the properties of 1088 Lyman Break Galaxies (LBGs)
 at $z\sim3$ selected from a $\sim2.63$ deg$^2$ sub-region of the
 First Look Survey field using the ground-based multi-color data 
 and the Spitzer Space Telescope mid-infrared data at 3--8 and 
 24 $\mu$m. With the wide area and the broad wavelength coverage, 
 we sample a large number of ``rare'' $u$-band dropouts which are 
 massive ($M_*>10^{11}M_{\odot}$), allowing us to perform a 
 statistical analysis of these subsets of LBGs that have not 
 been studied in detail. Optically bright ($R_{AB}\le24.5$ mag)
 LBGs detected in mid-infrared ($S_{3.6\mu m}\ge 6\mu$Jy) reside
 at the most massive and dusty end of the LBG population, with 
 relatively high and tight $M/L$ in rest-frame near-infrared. Most
 infrared-luminous LBGs ($S_{24\mu m}\ge100\mu$Jy) are dusty
 star-forming galaxies with star formation rates of 100--1000
 $M_{\odot}$/yr, total infrared luminosity of $>10^{12}L_{\odot}$.
 By constructing the UV luminosity function of massive LBGs,
 we estimate that the lower limit for the star formation rate
 density from LBGs more massive than $10^{11}M_{\odot}$
 at $z\sim3$ is $\ge3.3\times10^{-3}M_{\odot}/yr/Mpc^{3}$, 
 showing for the first time that the UV-bright population of 
 massive galaxies alone contributes significantly to the
 global star formation rate density at $z\sim3$.
 When combined with the star formation rate densities at $z<2$,
 our result reveals a steady increase in the contribution of massive
 galaxies to the global star formation from $z=0$ to $z\sim3$, 
 providing strong support to the downsizing of galaxy formation.
\end{abstract}


\section{INTRODUCTION}

 Stellar population analysis of local galaxies shows that
 stars in massive galaxies formed early within a short period   
 in the history of the universe, while low-mass galaxies went
 through late, slow star formation (Thomas et al. 2005;
 Heavens et al. 2004; Panter et al. 2006).
  The appearance of passively evolving early-type galaxies above
 $z\ge1$ (Cimatti et al. 2002; F\"orster-Schreiber et al. 2004)
 supports the idea of early star formation in massive systems.
  The evolution of specific star formation rate, i.e., the star
 formation rate per unit stellar mass shows that the star formation
 in most massive galaxies ($M_* \gtrsim 10^{11} ~M_{\odot}$) have nearly
 completed by $z\sim1.5$ while that for less massive galaxies continue
 to date (Papovich et al. 2006).
  These observations support the ``downsizing'' scenario of
 galaxy formation, commonly expressed as the decrease of stellar
 masses of galaxies in which vigorous star formation occurs with
 decreasing redshift
 (Cowie et al. 1996; Kodama et al. 2004; Juneau et al. 2005).
  
  Attempts have been made to explain this seemingly anti-hierarchical
 behavior of galaxy formation using hierarchical galaxy formation
 models. In an attempt to explain downsizing within the frame of 
 semi-analytic model, Neistein, van den Bosch, and Dekel (2006) 
 distinguishes ``archaeological downsizing ''(ADS) versus
 ``downsizing in time''(DST). In ADS, the downsizing is characterized
 as the built-up of stellar masses where massive galaxies are assembled
 through mergers of less massive galaxies which formed stars early.
 In comparison, DST is characterized by the formation and assembly
 of massive systems at early epoch. Semi-analytical models seem to be
 able to explain ADS naturally, while complicated baryonic processes
 are necessary to explain DST. Recent semi-analytic models have shown
 that the inclusion of AGN feedback effect successfully reproduce
 the observed decrease of actively star-forming system, yet more 
 understanding about the observed quantities and cooling processes are 
 needed. On the other hand, some other models do not seem to have much
 difficulty explaining the abundance of massive galaxies at high redshift
 (e.g., Nagamine et al. 2005). 

  Recent observational evidences point toward the DST-type of
 downsizing which suggests that there should be ``massive'', 
 star-forming galaxies at high redshift. Some of the red, massive
 galaxies are found to be dusty star-forming galaxies, and submm 
 galaxies are thought to be massive galaxies in formation at high 
 redshift (Smail et al. 2002). Still, it is not clear if they account
 for the whole star formation activity of massive galaxies. To complete
 the picture of the downsizing galaxy formation and to provide
 additional constraints to refine the models, we need to identify
 and study other actively star-forming massive galaxies, such as
 Lyman break galaxies (LBGs).

 Lyman break galaxies, galaxies selected by the continuum break
 at the Lyman limit, are the most commonly studied high-redshift
 star-forming galaxies. Since their selection technique requires 
 only optical imaging observation, many LBGs up to $z\sim6$ have 
 been discovered and studied to date (Steidel et al. 1999; 
 Papovich et al. 2001; Shapley et al. 2001; Giavalisco 2002; 
 Ouchi et al. 2004; Bouwens et al. 2004).
 These studies reveal that typical $z\sim3$ LBGs are galaxies
 with large ongoing star formation of 10--100 $M_{\odot}$/yr
 (Papovich et al. 2001; Shapley et al. 2001).
 Typical stellar masses of $z\sim3$ LBGs are found to be mostly 
 of order of 10$^{10}~M_{\odot}$ or less from the spectral
 energy distribution  fitting using optical and near-infrared photometry,
 although LBGs with masses up to $10^{11}~M_{\odot}$ were also found 
 (Papovich et al. 2001; Shapley et al. 2001).
 Some simulations suggest that nearly 50\% of $\ge10^{11} M_{\odot}$ 
 galaxies could be detected using the rest-frame UV selection criteria
 (Nagamine et al. 2005), but the observed fraction of LBGs among
 massive galaxies is less than 20\% (van Dokkum et al. 2006).

  Recently, the addition of \textit{Spitzer} mid-infrared (MIR)
 observations not only reduced the uncertainties in derivation
 of stellar mass (Shapley et al. 2005), but also enabled 
 the estimation of the infrared luminosities of LBGs.
 Huang et al.(2005) have defined Infrared Luminous LBGs (ILLBGs)
 as LBGs detected in the \textit{Spitzer} MIPS 24 $\mu$m with
 $f_{24\mu m} > 60\mu$Jy. The infrared
 luminosities of ILLBGs are estimated to be larger than
 $10^{12} L_{\odot}$,
 and they contribute $\sim5$\% of total LBG population. The stellar
 masses of ILLBGs exceed $5\times10^{10} M_{\odot}$ (Rigopoulou et al.
 2006), and their star formation rates inferred from infrared luminosity
 are as high as $\sim1000 M_{\odot}$/yr that is sufficient to evolve
 into present-day giant ellipticals.
 Several massive LBGs, not necessarily ILLBGs, are also identified
 in a part of the Extended Groth Strip covering $\sim227$ arcmin$^2$
 (Rigopoulou et al. 2006). These massive LBGs are found to have
 a star formation rate (SFR) of $\gtrsim100 M_{\odot}$/yr.
 These results suggest that ILLBGs and massive LBGs are important
 indicators of the star formation activity in massive galaxies
 at high redshift, along with the distant red galaxies
 (DRGs; Franx et al. 2003) and the submm galaxies.

 However, the number of $z\sim3$ massive LBGs and ILLBGs is still 
 small due to the difficulty of covering large areas with
 expensive $u$-band observations in order to discover such rare objects.
 For example, only 5 massive ($>10^{11} M_{\odot}$) LBGs are found
 in Rigopoulou et al.(2006). At $z\sim3$, the largest LBG sample
 (2347 photometric LBGs) until now is gathered from 11 separate fields
 of $\sim0.38$ deg$^2$ in total (Steidel et al. 2003).
 LBGs selected in the ESO Deep Public Survey (Hildebrandt et al. 2007)
 covers as much as $\sim1.75$ deg$^2$ at bright end
 ($30\arcmin\times30\arcmin\times7$; $R_{AB}\le24.0$), but until now, 
 the survey consists of optical data only. 

 In this study, we enlarge the $z\sim3$ massive LBG/ILLBG sample
 and investigate their properties using \textit{Spitzer} First Look Survey.
 The survey is not as deep as other multi-wavelength surveys, however,
 it is wide enough ($\sim4.3$ deg$^2$, of which 2.63 deg$^2$ was used
 in this study) to collect the rare bright LBGs.
 LBGs selected in optical bands (with $R$-band magnitude cut of
 $R_{AB}\le24.5$ mag) are combined with detections in
 \textit{Spitzer} IRAC (3.6--8$\mu$m) and MIPS (24$\mu$m).
 By comparing observed flux with model galaxy spectra, we estimate
 ages, the amount of dust extinction, and stellar masses of the LBGs.
 We also present 24$\mu$m detected LBGs in our study, and compare them
 with the ILLBGs in Huang et al.(2005).

  The data set we used is described in \S 2. We describe our selection
 criteria for $z\sim3$ LBG candidates in \S 3. The analysis on the stellar
 population of LBGs follows in \S 4, and the key results are presented
 in \S 5, such as the luminosity function, the stellar mass properties,
 and dust properties. We also discuss the implication of our 
 study on the galaxy evolution, in particular regarding the
 ``downsizing'' picture of galaxy formation in \S 5. Finally in \S 6,
 the conclusion follows. 
 Throughout this paper, we use a cosmology with
 $\Omega_{M}=0.3$, $\Omega_{\Lambda}=0.7$, and $H_0=70km/s/Mpc$.
 All the magnitudes are specified in AB system, unless noted.

   \begin{figure}
   \epsscale{1.3}
  \plotone{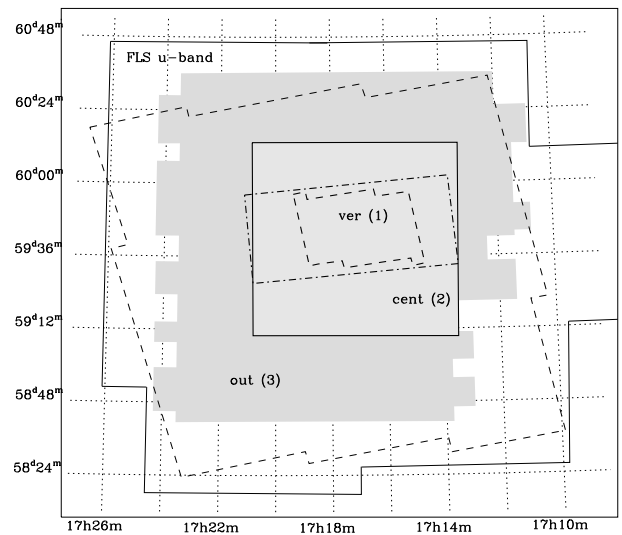}
  \caption{\label{fig:cov}
     The area coverage maps of various datasets in XFLS.
     The smallest region marked with thick\textit{dashed} line in the center
    represents verification field for IRAC ($\sim900$ arcmin$^2$)
    that is slightly deeper than other parts of XFLS. The skewed
    rectangle (\textit{dash-dot-dash} line) is the MIPS verification
    strip. The ``central 1 deg$^2$'' region is drawn as a thick square in the
    center. The shaded area outside the central 1 deg$^2$ region
    represents the outer field ($\sim1.63$ deg$^2$) where
    bright $u$-dropouts are selected.
     The IRAC main field of XFLS is specified with the thin \textit{dashed}
    line at outermost. Thin \textit{Solid} line shows our $u$-band coverage.
    }
  \end{figure}

\section{DATA}

  The LBG candidates are selected from the extragalactic component 
 of the \textit{Spitzer} First Look Survey (XFLS).
  XFLS comprises of InfraRed Array Camera
 (IRAC, 3.6, 4.5, 5.8, 8.0$\mu$m; Fazio et al. 2004) and
 Multiband Imaging Photometer for Spitzer (MIPS, 24, 70, 160$\mu$m;
 Rieke et al. 2004) imaging observations
 over the $\sim4.3$ deg$^2$ field centered at RA$=17^h 18^m 00^s$,
 DEC$=+59\arcdeg 30\arcmin 00\arcsec$.
 With an effective exposure time of 1 minute per pixel, the 
 main survey field ($\sim4.3$ deg$^2$) has 5 $\sigma$ flux limits of 
 20, 25, 100, and 100$\mu$Jy at wavelengths of IRAC 3.6, 4.5, 5.8,
 and 8.0$\mu$m respectively. The flux limit of MIPS 24$\mu$m
 is about $\sim300\mu$Jy (5 $\sigma$). The central part of the FLS
 ($\sim900$ arcmin$^2$) was observed deeper,
 with 10-minute integration times per pixel. 
 This ``verification'' field has sensitivities of 10, 10, 30, 
 and 30$\mu$Jy at IRAC wavelengths
 (5 $\sigma$; Lacy et al. 2005). For the MIPS verification field, 
 3 $\sigma$ flux limits are 90$\mu$Jy at 24$\mu$m (Fadda et al. 2006), 
 9 and 60 mJy at 70 and 160$\mu$m respectively (Frayer et al. 2006).  

  In order to select $u$-dropout objects in the XFLS,
 we used $u$, $g$, and $R$-band ground-based images acquired by
 three different wide-field instruments. We obtained
 deep $u^*$-band images with MegaCam on the Canada-France-Hawaii
 Telescope (CFHT) 3.6-m telescope (Shim et al. 2006).
 CFHT $u^*$-band filters are slightly redder than popular SDSS
 $u$-band. Therefore using the similar color selection criteria, 
 we expect that the $u^*$-band dropouts are biased towards 
 higher-redshift objects than previously selected LBGs.
  The central 1 deg$^2$ field was observed to the depth of
 $u^*\sim$26.2 mag (AB, 5$\sigma$ in 3$\arcsec$ diameter
 aperture).
 In addition to the central part, the whole FLS area was covered by
 $u^*$-band imaging with a shallower depth ($\sim$24.5 mag). Detailed
 information about the observations, photometry and the properties 
 of the dataset are presented in a separate paper (Shim et al. 2006).
  Deep $g$-band image was obtained together with the $u^*$-band
 observation over the central 1 deg$^2$ ($g\sim26.5$ mag).

  The $g^\prime$-band images were obtained over $\sim2.63$ deg$^2$,
 including the central 1 deg$^2$ area, using Large Format Camera
 (LFC; Simcoe et al. 2000) at the Palomar 5-m
 telescope (Storrie-Lombardi et al., in preparation). The LFC
 $g^\prime$-band images reach to the depth of 24.5 mag. For the region
 surrounding the central 1 deg$^2$ area, we use the LFC data for
 $g$-band photometry.
  The $u^*, g$ catalogs were extracted through dual-mode photometry of
 SExtractor (Bertin \& Arnouts 1996), using the $g$-band image as a 
 reference image.
  The $R$-band catalog of the FLS (Fadda et al. 2004)
 were matched with $u^*$, $g$-band catalogs using the matching
 radius of 0.8$\arcsec$, to obtain the $R$-band flux.
  The depth of the $R$-band image varies depending on the field
 location. Still, it reaches $R(Vega)\sim24.5$ mag at the deepest
 central field. Our method to select 
 $z\sim3$ LBGs with optical colors is described in detail in Section 3.

  Other ancillary datasets used to investigate the properties of
 $u$-dropout objects include $i^\prime$, $J$, and $K_s$-band images.
 LFC $i^\prime$-band images cover the same area with 
 the LFC $g^\prime$-band images with the
 depth of $\sim24$ mag. For the central part less than 1 deg$^2$,
 we also used NIR ($J(Vega)\sim21$ mag, $K_s(Vega)\sim20$--21 mag) data,
 obtained with WIRC on the Palomar 5-m ($K_s$-band), and FLAMINGOS on
 the KPNO 4-m ($J$-band; Choi et al., in preparation).
  The whole FLS field was covered by 1.4 GHz radio observations with
 VLA (flux limit at 90 $\mu$Jy, Condon et al. 2003). In this study,
 spectroscopic redshifts acquired with DEIMOS/Keck for 
 $\sim$1300 objects at $z\sim1$ (Choi et al. 2006)
 were used to check if there is any low-redshift interlopers within
 the LBG sample.

\section{LBG SAMPLE SELECTION}

  \subsection{Photometric Selection Criteria for $u$-dropouts}

  We select $u$-dropout objects from the 2.63 deg$^2$ area 
 that is covered by $u^*$, $g$, $R$, and $i$-bands.
 This area is illustrated as a shaded region in Figure \ref{fig:cov}.
 We define $u$-dropouts as objects that satisfy $R$-band magnitude 
 cut and color-cut criteria in $g-R$ versus $u^{*}-g$ diagram.
 The method is in principle identical to the popular ``drop-out''
 method for selecting LBGs (e.g., Steidel \& Hamilton, 1993).
 
  The adopted selection criteria can be summarized as follows:

\begin{center} 
   \begin{tabular}{l}

    $u-g\ge1.4$, \\
    $u-g\ge3(g-R)-0.12$, \\
    $g-R\le1.2$, \\
    $R \le 24.5$ (for central 1 deg$^2$), $R \le 23.5$ (for the outer area)
   \end{tabular}
\end{center}

  Different $R$-band magnitude cuts are adopted taking into account
 the difference in image depth (mainly $g^\prime$-band)
 between the central and the outer regions.
  For the central 1 deg$^2$ field, we apply a magnitude cut of
 $R\le24.5$ magnitude. For the outer 1.63 deg$^2$ area surrounding
 the central region, we adopt $R\le23.5$ magnitude cut.
  Note that $u$, $g$, and $R$ magnitudes are corrected for the
 galactic extinction (Shim et al. 2006; Fadda et al. 2004).

\begin{figure}
  \epsscale{1.2}
  \plotone{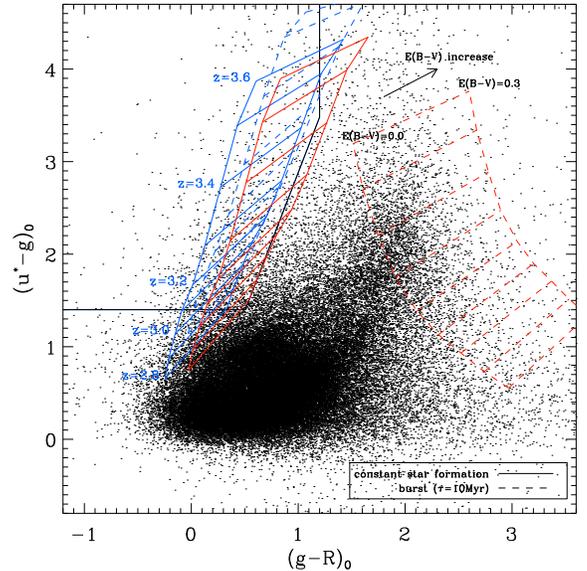}
  \caption{\label{fig:ccd}
   \small{
    The $g-R$ versus $u^{*}-g$ color-color diagram of galaxies
   in XFLS. Using the expected tracks of galaxies at the redshift
   range of $2.8<z<3.6$ with different age/star formation history/dust
   extinction, we identify the boxed region in the upper left as
   the place where the LBGs reside.
    The \textit{solid} lines indicate galaxies with constant
   star formation, and the \textit{dashed} lines indicate
   galaxies with single-burst star formation history.
   Different line colors indicate different ages, blue for 50 Myr-old
   galaxies and red for 1 Gyr-old galaxies.
    The shifts in x-axis direction represent the amount of dust extinction.
    Galaxies with constant star formation younger than 2 Gyr can be
   selected using the specified criteria. Young galaxies that have
   burst-like star formation history can also be selected, but old
   galaxies at their passively evolving stages (red, \textit{dashed}
   line on the right) cannot be selected with this criteria.}
   }
  \end{figure}

  The colors of star-forming galaxies at $2.8<z<3.6$ falls into
 the selection box (Figure \ref{fig:ccd}). The lines in the upper
 left of the color-color plot represent the location of model galaxies
 with constant star formation rate at $z\sim3$, of various ages and
 extinction values (see Figure \ref{fig:ccd} caption).
 Note that the lines represent galaxies with the solar metallicity,
 and the metallicity lower than the solar would
 move the lines to the left.

  \begin{deluxetable*}{ccccc}
  \tabletypesize{\small}
  \tablecaption{Surface Density of $u$-band Dropout Objects
   \label{tab:num_lbg}} 
   \tablehead{
    \colhead{$R_{AB}$ magnitude} &
    \colhead{N \tablenotemark{a}} &
    \colhead{number density \tablenotemark{b} } &
    \colhead{Steidel+99 \tablenotemark{c} } & 
    \colhead{Hildebrandt+05 \tablenotemark{d} }
    \\
    \colhead{} & \colhead{} & \colhead{(deg$^{-2}$)} &
    \colhead{(deg$^{-2}$)} & \colhead{(deg$^{-2}$)}
    }
   \startdata
      \tableline
      22.0--22.5    & 11  &  4.18$\pm$1.26 (1.72) &    \nodata     &  \nodata  \\
      22.5--23.0    & 35  & 13.31$\pm$2.25 (4.33) &  6.77$\pm$3.38  &  34.2$\pm$10.8 \\
      23.0--23.5    & 188 & 71.48$\pm$5.21 (20.6) &  71.06$\pm$13.54  & 72$\pm$18  \\
      23.5--24.0    & 247 &  247$\pm$15.7 (35.5)  &  294.4$\pm$37.22  & 234$\pm$28.8 \\
      24.0--24.5    & 607 &  607$\pm$24.6 (81.9)  &  656.50$\pm$54.14  & 540$\pm$54  \\
       \tableline
                     & 1088 &              &                  &
   \enddata
      \tablenotetext{a}{$u$-band dropouts are selected in the
      effective area of 2.63 deg$^2$ when $R\le23.5$ mag, and
      1 deg$^2$ when $R\le24.5$ mag.}
      \tablenotetext{b}{The errors are given for the Poisson noise only,
      and in the parenthesis, the errors considering both the Poisson
      noise and the galaxy clustering noise
      ($\sigma^2=\sigma_{Poisson}^2+\sigma_{clustering}^2$) are given.
      To calculate the clustering error, we used the approximation
      formula in Peebles (1975), adopting the amplitude of
      the angular two-point correlation function of $I_{AB} < 24.5$ LBGs
      in Foucaud et al.(2003). The value in Foucaud et al.(2003) was
      consistent with our own derivation within $\sim30$ \%. }
      \tablenotetext{c}{The surface number density of $z\sim3$ LBGs from
      Steidel et al.(1999). The errors given here represents
      the Poisson error.
       Steidel et al.(1999) used  $\mathcal{R}$ filter,
      which is slightly different from $R$ filter
      with $\mathcal{R}_{AB}-R_{AB}=-0.11$ (Foucaud et al. 2003).}
      \tablenotetext{d}{The surface number density of $z\sim3$ LBGs
      ($u$-dropouts) in CDF-South (Hildebrandt et al. 2005).
       The areas used in their studies are $\sim900$ arcmin$^2$.}
   \end{deluxetable*}

      \begin{deluxetable*}{lcccccc}
  \tablecaption{\label{tab:group} Subsets of LBG Sample}
  \tablewidth{0pc}
  \tabletypesize{\small}
  \tablehead{
  \colhead{Group} &
  \colhead{total} &
  \colhead{3.6$\mu$m} &
  \colhead{4.5$\mu$m}  & \colhead{5.8$\mu$m} &
  \colhead{8.0$\mu$m} & \colhead{24$\mu$m}
  }
  \startdata
    \tableline
    ver (1) ($R<$24.5)  & 189 & 22 & 21 & 6 & 7  & 4  \\
    cent(2) ($R<$24.5)  & 736 & 21 & 20 & 2 & 3  & 2  \\
    outer (3) ($R<$23.5)& 163 & 20 & 18 & 4 & 4  & 6
  \enddata
    \tablecomments
    { (1) The IRAC verification field ($\sim0.22$ deg$^2$)
      (2) The central part excluding the verification
     (i.e., effective area: $\sim0.77$ deg$^2$)
      (3) The outer $\sim1.63$ deg$^2$ field
     }
  \end{deluxetable*}

   The color selection criteria can select star-forming galaxies
 with constant star formation rate at the age of $\sim10$ Myr to
 $\sim2$ Gyr. The criteria can also select star-forming galaxies
 with exponentially decaying star formation rate at the age up
 to the exponential time scale $\tau$. Regardless of its star
 formation history, a galaxy younger than $\sim100$ Myr falls into
 the selection box. When there is internal extinction, the limit
 on age of galaxies that can be selected with the criteria changes.
 For example, the star-forming galaxies with age less than 500 Myr
 can be selected if $E(B-V)\lesssim0.3$.
  Passively evolving galaxies as old as $\sim1$ Gyr at $z\sim3$,
 which are growing old after a single burst, cannot be selected by
 this criteria (red, dashed line on the right in Figure \ref{fig:ccd}).

  After selecting $u$-dropouts from the color-color space,
 we remove spurious objects through the visual inspection.
  We find that spurious objects are mostly lying near
 the edge of the image, or too close to bright stars.
  On the other hand, we exclude objects that are likely to be a
 low-redshift interlopers based on either spectroscopic redshifts
 or photometric redshifts (see SED fitting section on Section 4.1).
 
  Finally as a result, we identify 1088 $u$-dropouts in our search
 area (925 objects with $R\le24.5$ mag in the central 1 deg$^2$,
 and 163 objects with $R\le23.5$ mag in the outer 1.63 deg$^2$).
 Table \ref{tab:num_lbg} shows the result.
  As mentioned in the above paragraph, the numbers in Table
 \ref{tab:num_lbg} are corrected for possible low-redshift interlopers
 based on their photometric redshifts or spectroscopic redshifts.
 At $R>23$ mag, the number density of $u$-dropouts of our study is
 consistent with the LBG surface density from previous studies
 (e.g., Steidel et al. 1999). At $22.5<R<23$, our surface density
 appears to be about twice higher than those found in Steidel et
 al.(1999), although 1-$\sigma$ error bars of the two numbers overlap.
  On the other hand, our number is
 about three times lower than those found by Hildebrandt et al. (2005).
 The possible explanations for the discrepancy are (i) low-redshift
 interlopers; (ii) the cosmic variance; and (iii) AGN contribution
 (see Section 5.1 and Figure \ref{fig:lf}). Spectroscopic observation of
 bright LBGs should be able to determine which explanation is right.
 Our study has an advantage over previous ones in terms of the
 area coverage ($\sim2.63$ deg$^2$), thus the Poisson errors
 in the number density estimates are significantly reduced.

   \begin{figure*}
  \plotone{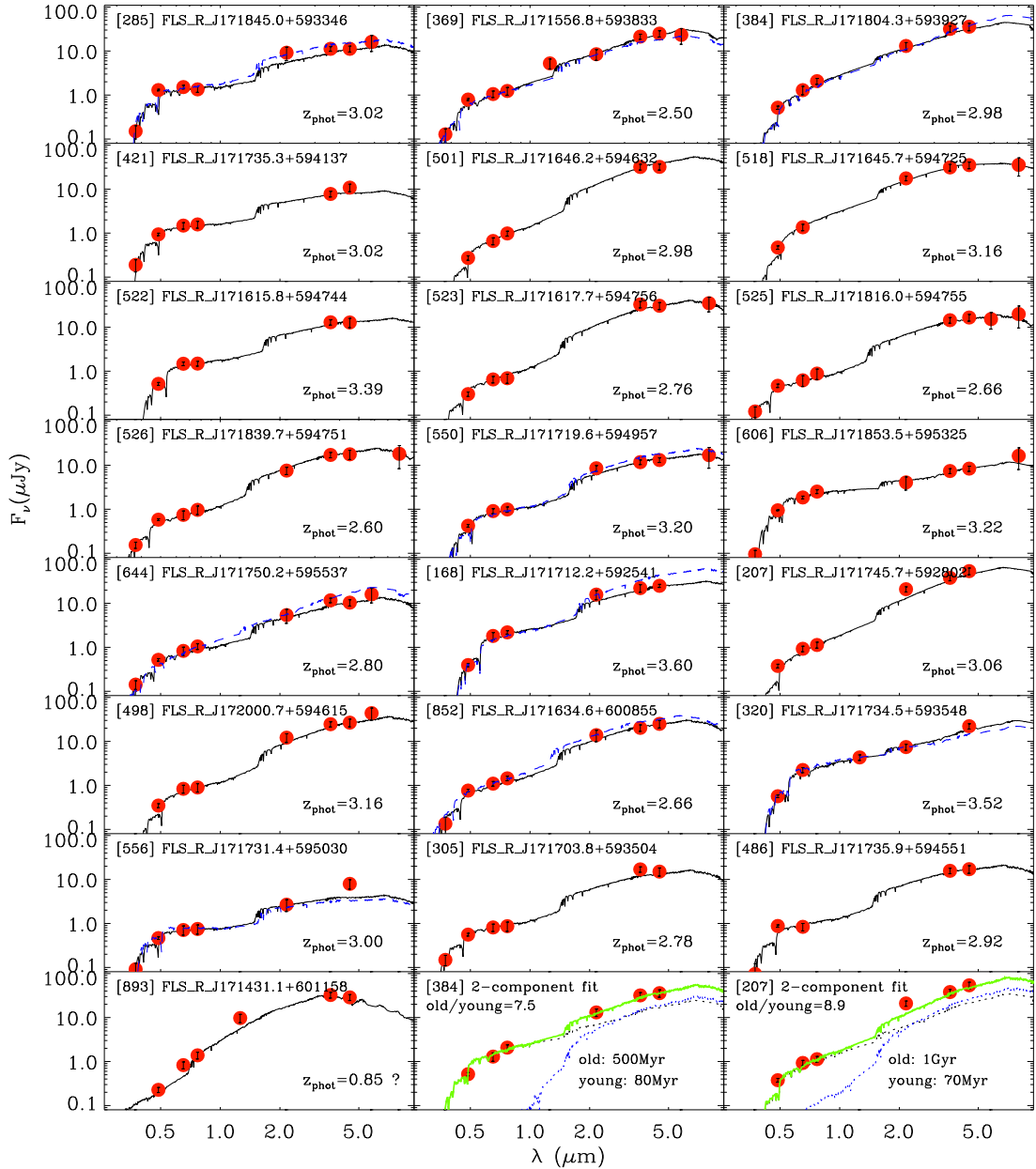}
  \caption{\label{fig:sed_ex}{
 The SED fitting results for a representative subset of our LBG sample.
 The x-axis is the observed
 wavelength, and the photometric data points are indicated for
 $u$, $g$, $R$, $i$, $J$, $K_s$, IRAC channel 1, 2, 3, and 4
 at 0.374, 0.487, 0.651, 0.768, 1.26, 2.16, 3.6, 4.5, 5.8, and 8.0
 $\mu$m. The object IDs are drawn from $R$-band catalogs of
 Fadda et al.(2004).
  The \textit{solid} line is the best-fit template using
 all the photometry data-points. For some objects,
 a \textit{dashed} line that indicates the best-fit
 template without MIR data-points is also overplotted.
 } }
 \end{figure*}

 \subsection{Mid-Infrared Detection of LBGs}

  The infrared fluxes of $u$-dropout objects are obtained from
 the published IRAC/MIPS source catalogs when available
 (Lacy et al. 2005; Fadda et al. 2006).
  If the object flux is not available, we measured the object flux
 through empirical PSF fitting, applying aperture correction values
 from Lacy et al.(2005).
  To reduce uncertainties in flux measurement, we use only objects
 whose fluxes exceed 3-$\sigma$ flux limits in IRAC.
  The number of $u$-dropouts detected in each IRAC band
 is summarized in Table \ref{tab:group}. Here, we define LBGs
 detected over 3-$\sigma$ limit in 3.6 $\mu$m image as an
 IRAC-detected LBGs (hereafter, IRAC LBGs) since all the objects
 detected in other IRAC bands are also detected in 3.6 $\mu$m.

  The number of IRAC LBGs is dependent on the depth of the Spitzer
 image. For example, the number of detections in Table \ref{tab:group}
 is relatively small compared to the deeper surveys because of
 the shallower depth of our survey (e.g., Barmby et al. 2004; 
 Huang et al. 2005).
  Note that, at the same infrared flux limit, the number of
 IRAC detections in our study is consistent with the other
 studies. For example, we find 22 IRAC LBGs above the
 3 $\sigma$ flux limit of 6 $\mu$Jy (3.6$\mu$m) in the FLS
 verification strip (900 arcmin$^2$).
  On the other hand, Huang et al.(2005) have identified
 8 LBGs and 3 QSO/AGNs in 3.6 $\mu$m image at the same flux
 limit, over 227 arcmin$^2$ (see Figure 1 in Huang et al. 2005).
 If we scale the area coverage, we would expect that we could find
 $\sim6$ objects in 227 arcmin$^2$ of Huang et al.(2005). These
 numbers are relatively consistent with each other.
  Huang et al.(2005)'s sample includes LBGs as faint as
 $24.5 < R < 25.5$ while we do not. This $R$-band difference
 may cause a slight discrepancy between the two numbers, but
 the difference is almost negligible for the LBGs bright in
 the rest-frame NIR wavelengths which also turn out to be
 massive LBGs. Given the bright rest-frame NIR flux, we find that
 IRAC LBGs belong to massive, IR-bright end of LBGs (see Section
 5.2 for more detail).
  Our sample is biased against UV-faint LBGs due to the bright
 $R$-band magnitude cut, but the above comparison suggests that 
 the bias does not affect our analysis of massive LBGs since
 they are mostly bright in IRAC wavelengths.

  For the MIPS detection, we matched our IRAC LBGs with
 MIPS 24 $\mu$m sources in Fadda et al.(2006).
 In the verification field, 6 IRAC LBGs found matches in the
 24 $\mu$m catalog (over 5 $\sigma$ detection, corresponding to
 $S_{24}\sim150\mu$Jy). From the shallower parts of the FLS field,
 6 objects are found to be in the 24 $\mu$m catalog
 (5 $\sigma$, 300$\mu$Jy).
 We call these LBGs detected in 24 $\mu$m as ``24 $\mu$m LBGs''
 \footnote{Among non-IRAC LBGs, 2 objects found their matches
 in 24 $\mu$m images. However, we didn't include them in 
 24 $\mu$m LBG sample since their properties are hard to constrain
 with limited optical photometry.}. 
  In total, we have twelve 24 $\mu$m LBGs from 1088
 $u$-dropouts in our sample. Similarly to the case of IRAC LBGs,
 MIPS-detection is dependent on the depth of the image, and
 the number of 24 $\mu$m detections can certainly go up
 if we had a deeper MIPS image.
 The 24 $\mu$m LBGs are IR-luminous sub-population of $z\sim3$
 IRAC LBGs with $L_{IR} > 10^{12} L_{\odot}$
 (See section 5.3 for more detail).

 \section{SED FITTING}

  \begin{figure*}
 \plotone{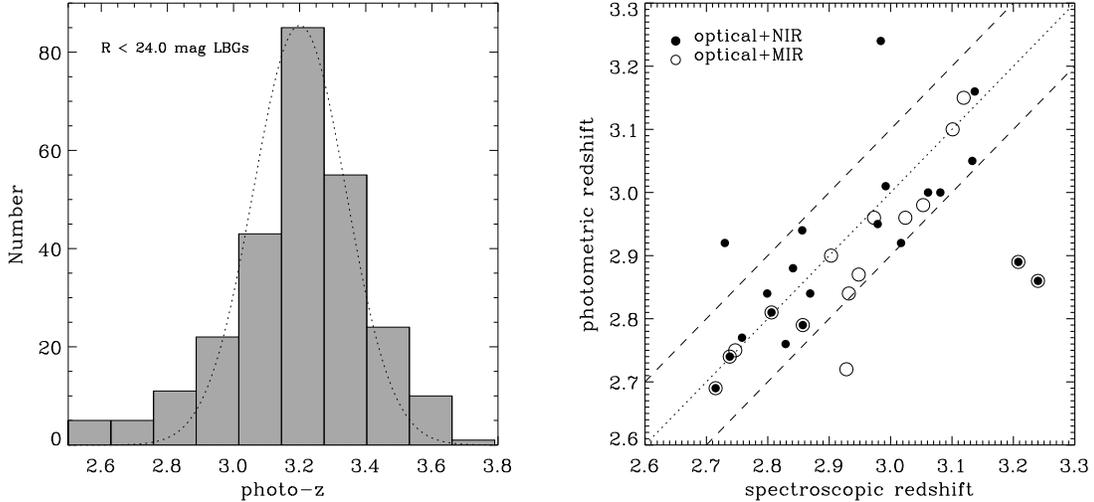}
 \caption{\label{fig:method_ver1}
 \textit{Left}: The photometric redshift distribution of our LBG
 sample derived from SED fitting. \textit{Dotted} line is an
 approximation of the distribution as a Gaussian function. Mean
 redshift $\langle z \rangle$ is $\sim3.2$, and the standard
 deviation of the distribution is $\sim0.14$.
 \textit{Right}: The comparison between spectroscopic redshifts and
 photometric redshifts of LBGs in the Westphal field
 (Steidel et al. 2003).
 Open circles represent 8 micron detected LBGs from Rigopoulou et al.
 (2006), while filled circles indicate other LBGs with at least
 one NIR ($J, K_s$) photometry. Among the 8 micron LBG sample,
 only 6 objects have NIR photometry (points with both filled and
 open circles).
 }
 \end{figure*}

  We have performed SED fitting in order to derive various
 parameters, especially focusing on the stellar masses of LBGs.
 The SED fitting method employs the same methodology as previous
 studies (e.g., Papovich et al. 2001; Shapley et al. 2001). Various
 galaxy SED templates were generated using a stellar population
 synthesis model (Bruzual \& Charlot 2003), and fitted to the
 observed SED.
  We used all the available photometric data points from $u$-band
 to IRAC 8 $\mu$m. Note that if we have $u, g$, and $R$-band data
 only, we did not perform the SED fitting since it was very difficult
 to gain meaningful fit-results from three photometric data points.
  For optical fluxes, we used MAG\_AUTO in SExtractor,
 while infrared fluxes were measured from the PSF fitting 
 or drawn from the catalog (Lacy et al. 2005) as 
 described in the previous section. 
  In the NIR, the LBG candidates were matched with $J$- and
 $K_s$-band catalogs (MAG\_AUTO from SExtractor; 
 Choi et al. 2007, in preparation) 
 within the matching radius of 1.2$\arcsec$. 

 For the generation of various galaxy SEDs, we limited our templates
 to a specific star formation history. The galaxies have the
 exponentially decaying form of ($\phi(t) \propto exp(-t/ \tau)$)
 star formation with $\tau$=10Myr, 100Myr, 300Myr
 or a constant star formation rate.
 In our case, the best-fit SED of almost all the LBGs gives a constant
 star formation template (cf. Rigopoulou et al. 2006).
  The metallicity and the initial mass function are fixed to certain
 values/form. The adopted metallicities are either 0.2 $Z_{\odot}$
 or 1 $Z_{\odot}$.
  The IMF was fixed as the Salpeter IMF between
 0.1 $M_{\odot}$ and 100 $M_{\odot}$.
  The stellar population was evolved using Padova 1994 stellar
 evolutionary tracks. The ages of a galaxy, $t$, was chosen from $\sim$22
 possible values between 10 Myr to the age of the universe
 at the corresponding redshift.
  The galactic reddening of the generated spectra is
 following the dust extinction law for starburst galaxy
 (Calzetti et al. 2000).
 We also added the flux suppression shortward of Ly $\alpha$
 forest due to the intergalactic medium following Madau et al.(1995).

  The best-fit parameters - photometric redshift $z$, extinction
 parameter $E(B-V)$, stellar mass, age, star formation history and
 metallicity - are obtained by minimizing the error-weighted
 $\chi^{2}$ value expressed as
 $\chi^2=\sum_{i, filter}\frac{(f_{obs}-\langle F_{\nu}\rangle)^2}{\sigma_{obs}^2}$.
 Note that ages, star formation histories, and metallicities are 
 chosen from discrete values as explained above. 
   
  An illustration of SED fitting for some galaxies is presented
 in Figure \ref{fig:sed_ex}, including SEDs of possible low-redshift
 interlopers.

   \begin{figure*}
 \plotone{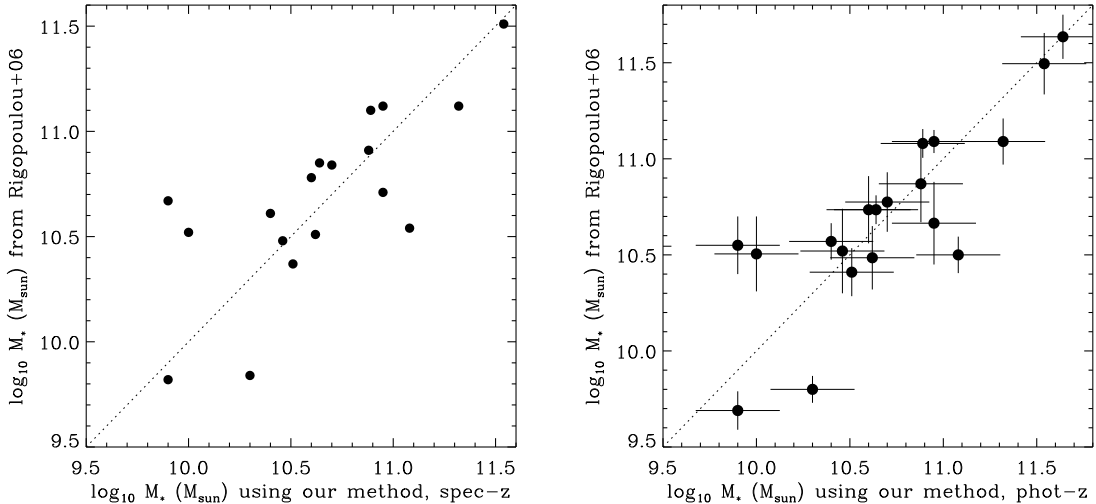}
 \caption{\label{fig:method_ver2}
 \textit{Left}: The comparison of the stellar masses
 derived using spectroscopic redshifts and our method against that of
 Rigopoulou et al.(2006).
 \textit{Right}: The comparison of the stellar masses using photometric
 redshifts and that of Rigopoulou et al.(2006).
 The stellar masses estimated by two different methods agree
 with each other within the errors.
 }
 \end{figure*}

 \subsection{Photometric Redshift}

 The distribution of photometric redshifts of $u$-dropouts is
 presented in Figure \ref{fig:method_ver1}a. The median redshift 
 of LBGs is $\langle z \rangle \sim3.2$, and the standard deviation 
 of the photometric redshift distribution of LBGs is $\sim0.14$.
 The $u$-dropouts are distributed mostly at $2.5<z<3.8$,
 but there is a low redshift tail to the distribution.
 We consider those having low photometric redshifts to be interlopers.

 For example, 5 objects out of the ``initial'' IRAC LBGs are turned
 out to have probable redshifts much less than 3. The best-fit SED 
 results for these objects are heavily attenuated, moderately old 
 galaxies at $z\sim1$ (see the first object in the last row of
 Figure \ref{fig:sed_ex} as an example).
 Among these, 3 objects were identified to lie at
 $z=0.46, 0.96$, and 1.26 in match with the spectroscopic sample of
 $z\sim1$ galaxies (Choi et al. 2006).
 The ``possible contaminants'' are characterized by
 red $g-R$ and $R-i$ colors, bright NIR flux ($J$), and decreasing
 MIR flux from 3.6 $\mu$m to 8.0 $\mu$m which suggests that
 1.6 $\mu$m H$^-$ bump has not redshifted out of 3.6 $\mu$m.
  For LBG candidates with only $g, R,$ and $i$ detections, 
 low-redshift interlopers are also identified as objects with
 red $g-R$ and $R-i$ colors.
  When there are even no $i$ detection, it was difficult to
 determine which $u$-dropouts are low-redshift interlopers.
 However, the non-detection in $i$-band suggests that they
 have blue $g-R$ colors which is indicative of LBGs at $z\sim3$.

  Through these breakdown of the low-redshift interlopers, we
 removed $u$-dropouts that have most probable photometric redshifts
 of $z<2$.
  Our investigation shows that the fraction of possible low-redshift
 interlopers is higher at the brighter $R$-band magnitude bin. 
  Among $u$-dropouts, the
 fraction of possible interlopers with low photometric redshifts
 is $\sim25$\% at the brightest ($22<R<22.5$)
 bin, $\sim13$\% at the second brightest ($22.5<R<23.0$) bin. For the 
 other bins, the fraction is of order of $\sim5$\%. 
  Three spectroscopically confirmed low-redshift galaxies 
 mentioned above and galaxies with low photometric 
 redshifts are excluded in Table \ref{tab:num_lbg} and 
 in the statistical analysis hereafter.
  A future spectroscopic study of bright LBGs will be able to 
 determine how reliable the estimates of the number of 
 interlopers are. 
 
 In order to check the reliability of our photometric redshift,
 we tested our method on spectroscopically confirmed LBGs
 in the extended Westphal-Groth strip (Steidel et al. 2003; hereafter 
 Westphal LBGs). For the photometric data, we used 
 the optical, NIR, and MIR photometry of Westphal LBGs in
 Shapley et al.(2001) and  Rigopoulou et al.(2006).
 All the LBGs used in this test have at least one photometric 
 data point at wavelength longward of the optical (i.e., NIR or MIR).

 Figure \ref{fig:method_ver1}b shows the comparison of our 
 photometric redshifts versus spectroscopic redshifts.
 Although there exist a few outliers, in overall the plot shows that 
 the derived photometric redshift is quite reliable 
 to about $\Delta z/z \sim0.1$, regardless of whether we have
 NIR or MIR data.

 \subsection{Stellar Mass}

 Stellar mass is known to be a robust parameter that can be 
 constrained relatively easily compared to other parameters,
 by being insensitive to the assumed star formation history
 (e.g., Papovich et al. 2001; Rigopoulou et al. 2006).
 Other constraints, such as metallicity, are found to affect
 the derived stellar mass within a factor of 2--5 (Papovich et al.
 2001). In addition to that, it has been reported that
 the inclusion of MIR photometry data points reduces the
 stellar mass uncertainties by a factor of 1.5--2 (Shapley et al. 2005).
 Here, we investigate how our stellar masses fare with the values
 from other methods, and also how much the lack of
 spectroscopic redshifts influences the stellar mass derivation.
 Again, we use the Westphal LBGs for this purpose (Rigopoulou et al. 2006).

 To inspect the uncertainty arising from photometric redshift,
 we performed this comparison using spectroscopic redshifts first
 (Figure \ref{fig:method_ver2}a). Then, we estimated stellar masses 
 using photometric redshifts instead of spectroscopic redshifts 
 (Figure \ref{fig:method_ver2}b), and studied how the use of the 
 photometric redshifts affects the result.

 Figure \ref{fig:method_ver2}a shows that
 our stellar masses derived with spectroscopic redshifts 
 are consistent with those from Rigopoulou et al.
 (2006) within $\le0.2$ dex. Therefore, we consider 
 the inherent amount of stellar mass uncertainty due to the fitting
 method to be $\sim 0.2$ dex. Figure \ref{fig:method_ver2}b shows
 that stellar masses derived with photometric redshifts are within 
 0.1--0.2 dex from those derived with spectroscopic redshifts.
  Considering that the uncertainty of stellar masses is dependent on
 the fitting method by about $\sim0.2$ dex, the stellar masses 
 derived with photometric redshifts and spectroscopic redshifts are
 consistent within the error from the fitting method.
  Two objects show relatively large discrepant values 
 (objects at around [10.0, 10.5]), but this can be explained with
 a failure of the fitting since these objects have
 large $\chi^2$ values.
 In addition, we have tested two-component fitting for several objects
 (see last 2 objects in Figure \ref{fig:sed_ex}) to see how the 
 presence of an underlying old stellar population 
 affects the derived stellar mass. 
  For the two components, we used an old component 
 (500Myr or 1Gyr passively evolving population after single burst)
 and a young component (constant star forming population
 younger than 100Myr). The result shows 
 $\sim20$\% of difference in the derived stellar masses compared to 
 those derived using a single-component fitting.
 Since the discrepancy is relatively small, we used single-component 
 fitting only.

 \section{RESULTS}

 \subsection{Bright End of the Luminosity Function}

 We have constructed the UV luminosity function of our LBGs
 using the entire sample corrected for interlopers 
 (Table \ref{tab:num_lbg}). At $2.5<z<3.8$ where our LBGs are
 distributed, the central wavelength of $R$-band corresponds to
 1360$\mbox{\AA}$--1860$\mbox{\AA}$. Therefore,
 we converted $R$-band magnitude to the absolute magnitude with
 the K-correction consisting of the bandwidth dilation term
 only, and used it as the UV absolute magnitude. The UV luminosity
 derived this way samples different rest-frame UV wavelengths at
 different redshift, but the effect due to this should be less 
 than a few tenths of magnitude.

 To derive the luminosity function, we used the $\frac{1}{V_{max}}$ 
 method (Schmidt 1968; Lilly et al. 1995; Im et al. 2002).
 The number density of LBGs in each magnitude bin is
 calculated as  
 \[ \phi(M) [mag^{-1} Mpc^{-3}] = \frac{1}{{\Delta}M}
 \times \sum \frac{1}{V_{max}} \]
 while $\Delta M$ is the size of the magnitude bin.
 The maximum comoving volume, $V_{max}$ is calculated by the equation
 \[ V_{max}=\int_{max(z_1,z_{min})}^{min(z_2,z_{max})} \frac{dV}{dz} dz \]
 where $z_1$ and $z_2$ indicate the lower and the upper limit of
 the redshift distribution of the galaxies.
 The $z_{min}, z_{max}$ values
 represent the minimum and maximum redshift where the galaxy can be
 detected. $dV/dz$ is the differential comoving volume at redshift $z$.

  \begin{figure}
  \epsscale{1.2}
  \plotone{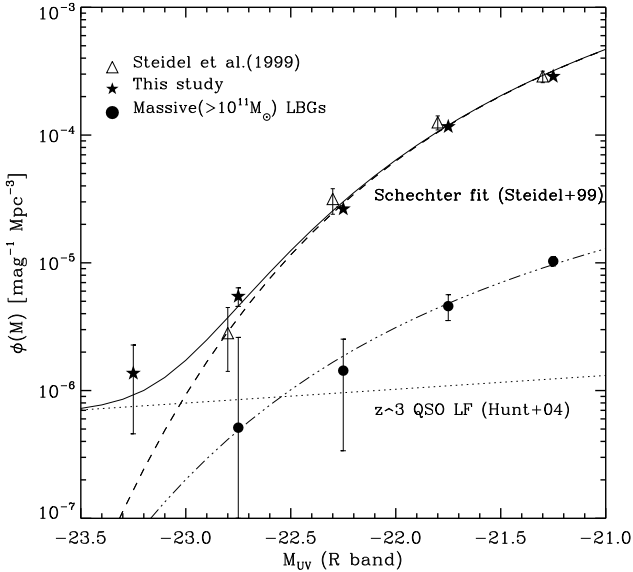}
  \caption{\label{fig:lf}
   The FUV luminosity function of our LBGs, compared with the LF of
  LBGs and QSOs from Steidel et al.(1999) and Hunt et al.(2004).
   Note that the size of error-bar of our LF, compared to previous
  results, has shrunken significantly at bright magnitudes
  due to the wide area coverage of our survey.
  \textit{Dashed} line is the best-fit Schechter function for $z\sim3$
  LBGs from Steidel et al.(1999), with $\alpha=-1.6$,
  $M_{*}=-21.04$ mag. \textit{Dotted} line is $z\sim3$ QSO luminosity
  function from Hunt et al.(2004).
  \textit{Solid} line is the sum of the $z\sim3$ LBG luminosity
  function and the $z\sim3$ QSO luminosity function. Our data points drawn
  with stars are consistent with the solid line.
  At $M_{UV}<-23.0$ mag, the QSO number
  density exceeds that of LBGs. The significant excess of LBGs at bright
  end over the Schechter function is most likely to be due to QSOs.
   The number density of massive ($>10^{11} M_{\odot}$) LBGs
  are overplotted as filled circles.
   We derived the best-fit parameters for Schechter function to
  be $\alpha=-1.6$, $M_{*}=-21.6$ mag (\textit{3-dot-dashed} line).
  }
  \end{figure}

 Our result is consistent with previous results 
 (Steidel et al. 1999) at the magnitude range of
 $-22.5< M_{UV}< -21.0$ (Figure \ref{fig:lf}).
 At the bright end, there are two noticeable results: (i) the decrease
 of error-bars, (ii) a clear excess of bright LBGs compared to the
 expected number from the best-fit Schechter function. First is due
 to the large area coverage of this study, since the error-bars in
 Figure \ref{fig:lf} represent the Poisson error only.
 When the galaxy clustering is taken into account, the error bar
 increases by up to a factor of 2 over the Poisson statistics.
 Even after the clustering effects are taken into account (Peebles 1975;
 please refer to the clustering error in Table \ref{tab:num_lbg}),
 our error bars are smaller than those of previous studies.

 The excess of bright LBGs are quite interesting, although it has
 been expected from the high surface density of LBGs at the
 bright $R$-band magnitude bin (Table \ref{tab:num_lbg}).
 The apparent $R$-band magnitudes of LBGs with $M_{UV}\le-23.0$ mag
 are all between $22\le R\le23$. Until now, few LBGs have been 
 discovered over $M_{UV}=-23.0$ mag, including six ``very'' bright
 ($M_{UV}\sim-25.4$ mag) $u$-dropouts discovered in
 SDSS DR1 ($\sim$1360 deg$^2$; Bentz et al. 2004).
 Possible candidates of these bright LBGs are galaxies magnified by
 gravitational lens (e.g., cB58; Williams \& Lewis 1996), quasars,
 late-type stars, and low-redshift interlopers as we noted in
 Section 4.1.
  Among these possibilities, the 
 most probable cause for bright-end excess is QSOs at $z\sim3$.
  According to a previous study of $z\sim3$ QSOs
 (Hunt et al. 2004), the number density of faint QSOs exceeds
 that of the galaxies at the magnitude range of $M_{UV}\le -23.0$
 (Figure \ref{fig:lf}).
  The comparison of our LBG luminosity function with the QSO
 luminosity function at $z \sim 3$ suggests that most of the
 excess above $M_{UV}\le -23.0$ can be explained by QSO.

  Only one of these $M_{UV}\le-23.0$ mag UV-luminous LBGs are
 detected in 24 $\mu$m image.  This may be due to the depth of 
 24 $\mu$m image, and does not constrain the possibility of $R$-band
 bright LBGs being AGNs. If AGN-like LBGs are included in stellar
 mass analysis, the presence of AGNs may boost up the derived stellar
 masses. Still, the stellar masses of these bright LBGs are
 $\sim10^{10} M_{\odot}$, with the most massive one of
 $2.6\times10^{10} M_{\odot}$. Therefore, the excess of these
 ``UV bright'' LBGs does not affect the analysis of ``massive'',
 i.e., $> 10^{11} M_{\odot}$ LBGs given in this paper.

 \subsection{Properties of IRAC LBGs}

  In this section, we investigate the properties of 63 IRAC LBGs 
 (detected in IRAC, $S_{3.6\mu m}>6\mu$Jy in the verification strip,
 $S_{3.6\mu m}>12\mu$Jy in the FLS main field). We focus on the
 stellar masses and dust properties of IRAC LBGs derived through
 SED fitting. 

 \subsubsection{$M/L$ ratio in rest-frame NIR}

  \begin{figure*}
  \plottwo{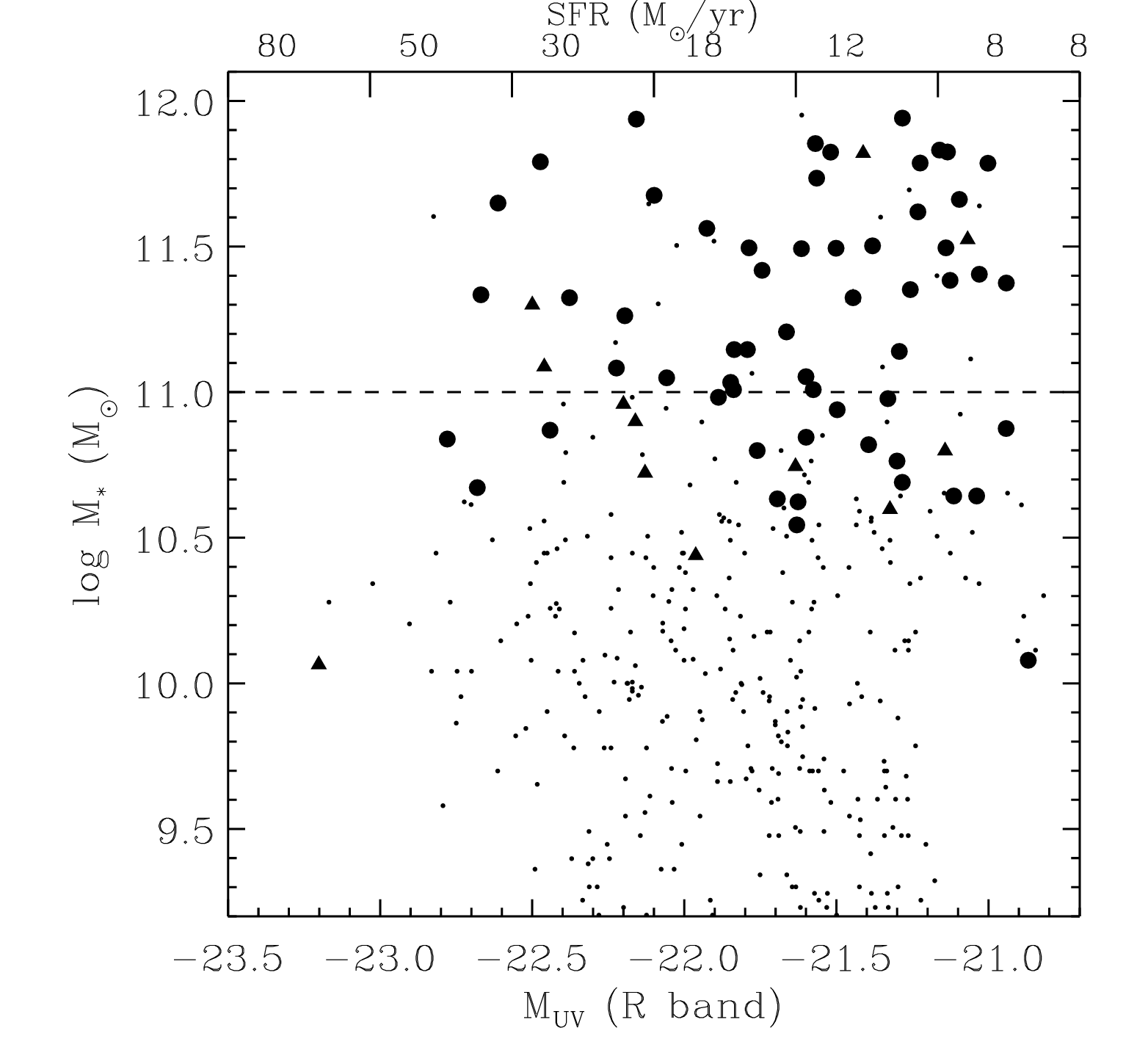}{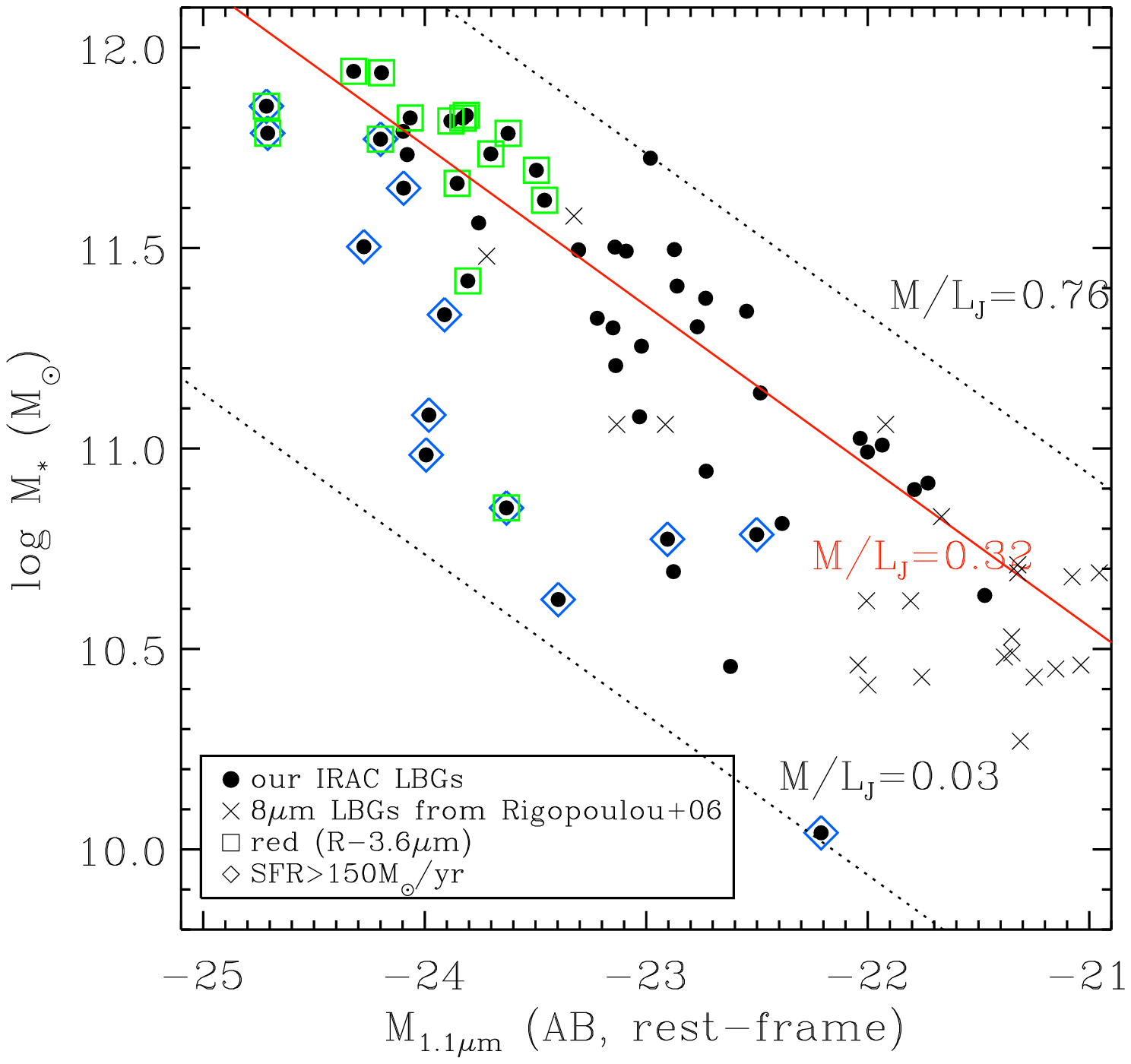}
  \caption{ \label{fig:mass_relation}
  The relation between stellar masses and magnitudes
  (\textit{left}: $R$-band, \textit{right}: 4.5 $\mu$m).
  The plot shows that there is a good correlation between
  stellar masses and $4.5\mu$m magnitudes (rest-frame $J$-band
  at $z\sim3$), while little correlation is shown in
  stellar masses vs. $R$-band magnitudes (rest-frame UV).
  (\textit{Left}): Filled circles for IRAC LBGs, filled triangles
  for NIR-detected LBGs with no detection in IRAC,
  and small dots for LBGs with optical data points only.
  (\textit{Right}): Filled circle are IRAC LBGs that have
  4.5$\mu$m detection. Galaxies with red optical-MIR color
  ($(R-3.6\mu m)>3.5$) are marked as open squares, and galaxies
  with ongoing star formation rate larger than 150 $M_{\odot}/yr$
  are specified with open diamonds. The correlation coefficient
  decreases to $r=-0.85$ from $r=-0.72$ when open diamonds are
  excluded. The dotted line demonstrates the range of
  stellar $M/L$ in solar unit. There is a sequence of IRAC LBGs
  of $M/L_{J}\sim0.32$, with an rms error of 0.16
  when objects with $M/L_{J} < 0.1$ are excluded.
  The stellar masses and $4.5\mu$m magnitudes of 8 micron LBGs
  from Rigopoulou et al.(2006) are overplotted as crosses.
  8 micron LBGs share similar $M/L$ properties with IRAC LBGs.
  }
  \end{figure*}
 
  Among 63 IRAC LBGs, 43 are identified to be more massive than 
 $10^{11} M_{\odot}$. The bright rest-frame NIR flux of IRAC LBGs
 suggest that these reside at the most massive end of the whole 
 LBG population, and Figure \ref{fig:mass_relation}a shows that 
 on average, IRAC LBGs are at the massive end of the mass 
 distribution of LBGs. In Figure \ref{fig:mass_relation}a, 
 filled circles represent IRAC LBGs, filled triangles represent
 LBGs detected in NIR but not in IRAC (13), and small dots represent
 LBGs detected in $g, R,$ and  $i$-bands only (501). 
 The range of the stellar masses of our LBGs is
 $2\times10^9 M_{\odot} < M_{*} < 10^{12} M_{\odot} $.
 There is no significant correlation between $R$-band magnitudes
 and stellar masses. We estimate the uncertainty involved 
 in stellar mass of each LBG to be of a factor of 2--3.

  In the local universe, the rest-frame NIR photometry can be
 a useful stellar mass indicator (e.g., Bell \& de Jong 2001).
  Even for high-redshift LBGs, the same rule applies. For example,
 Rigopoulou et al.(2006) have found a correlation between LBG
 stellar mass and the magnitude getting tighter as the wavelength
 of the band getting longer. In other words, $M/L$ ratio of galaxies
 derived in the longer wavelength has smaller scatter compared to
 the value derived in the shorter wavelength. They used LBGs with
 $M_{*}\lesssim10^{11} M_{\odot}$, and we extend the $M/L$ analysis
 of LBGs to objects more massive than $10^{11} M_{\odot}$, using
 rest-frame NIR wavelength.
 Since the number of 8 $\mu$m detected LBGs is small in our 
 sample, we use 4.5 $\mu$m (rest-frame $1.1\mu$m) flux here 
 instead of 8.0 $\mu$m flux (rest-frame $K$-band).
  Figure \ref{fig:mass_relation}b shows the 4.5 $\mu$m magnitude
 (the rest-frame $J$-band at $z\sim 3$) versus stellar mass.
  Our result shows that even above $10^{11} M_{\odot}$, there
 is a good correlation between the rest-frame NIR flux and
 the stellar mass, while no trend is found between the $R$-band
 magnitude and stellar mass (Figure \ref{fig:mass_relation}a).

  Despite the probable consistency of $M/L$ ratio at rest-frame NIR 
 wavelength, Shapley et al.(2005) reported that there is still 
 a scatter of $\sim10$ for $M/L$ ratio of star-forming galaxies
 at $z\sim2$ when measured in rest-frame 1.4 $\mu$m.
  We also find that there is an order of magnitude spread in
 $M/L$ ratio at 4.5 $\mu$m.
 The variation in $M/L$ ratio is mainly due to the differences in 
 star formation history. Old galaxies 
 have higher $M/L$ value in NIR, while young galaxies show lower $M/L$.
 We examined the $M/L$ ratios of model galaxy templates with different
 age and star formation history, and found that $M/L$ ratio of 
 1 Gyr-old galaxy with constant star formation rate is about 
 12 times smaller than 1Gyr-old galaxy that have been evolved
 passively after a single burst ($\tau=10$Myr). Comparing 
 passively evolving population of different age, the $M/L$
 ratio of galaxy can be more than twice higher when its age
 increases from 200 Myr to 1.5 Gyr.

 To demonstrate the above point, in Figure \ref{fig:mass_relation}b,
 we mark objects with red optical-MIR color of $(R-3.6\mu m) > 3.5$ 
 which are presumably old galaxies with squares and 
 those with high ongoing star formation ($>150 M_{\odot}/yr$)
 with open diamonds. Here, the star formation rates are derived from
 UV luminosity, corrected for the dust extinction using $E(B-V)$ obtained
 with the SED-fitting procedure. 
 The figure shows that actively star-forming galaxies 
 have $M/L$ ratios about 10 times smaller than red galaxies which have much 
 smaller scatter in the $M/L$. The star formation activity is a major 
 source that provides a spread of a factor of 10 in the mass-to-light
 relation even at IRAC wavelengths.
 The correlation coefficient between $M_{4.5\mu m}$
 and $log M_{*}$ is $-0.72$ for all LBGs in the
 figure. When LBGs with large SFR ($\ge 150 M_{\odot}/yr$) are
 excluded, the correlation coefficient is $-0.85$. 
  Also, when objects are restricted to galaxies with red (optical-MIR) 
 colors, we find a tighter correlation between stellar mass and the 
 rest-frame NIR luminosity.  Similar effect has also been 
 addressed in Shapley et al.(2005), that $(R-K)>3.5$ galaxies show
 tighter correlation in mass-to-light relation among $z\sim2$
 star-forming galaxies.

 \subsubsection{Comparison of IRAC LBGs and DRGs}

 Previous studies underline the exclusive characteristics among 
 the high-redshift galaxies selected by different criteria. For 
 example, LBGs and DRGs ($(J-Ks)>2.3$; Franx et al. 2003) are 
 thought to represent blue and small/red and large systems at high 
 redshift (van Dokkum et al. 2006). Little ($<10$\%) overlap between 
 the two galaxy populations is addressed (Labb\'e et al. 2005; van
 Dokkum et al. 2006). Since IRAC LBGs are at the massive end of the
 LBGs and some of them have red (optical-MIR) color, it is noteworthy
 to discuss whether these ``IRAC LBGs'' can also be selected with DRG
 selection criteria; or at least, whether their mass/color ranges are
 comparable with those of DRGs. 

  As for the color, the selection cut for DRGs is $(J-K)>2.3$, 
 which is sensitive to the underlying old stellar population. 
 There are only small number of IRAC LBGs with both $J$-and $K_s$-band
 detection, and their colors are between $0.5<(J-K_s)<1.6$. With 
 these, the IRAC LBGs cannot be selected by DRG criteria. However,
 red (optical-MIR) color cut, for example $(R-3.6\mu m)>3.5$
 to specify galaxies showing tight mass-to-light correlation, is 
 comparable with the color cut of DRGs. Considering the number of 
 $(R-3.6\mu m)>3.5$ LBGs, we estimate that at least 20\% of 
 IRAC LBGs are candidates that could be selected with DRG criteria. 

 Labb\'e et al.(2005) suggest that LBGs and DRGs are well-separated
 in $(I-K_s)$ vs. $(K_s-4.5\mu m)$ diagram in Figure 1 of their paper.
 LBGs have blue $(I-K_s)$ values ($0<(I-K_s)<2$), and DRGs have 
 $(I-K_s)>2$, redder $(K_s-4.5\mu m)$ color than LBGs. For the IRAC 
 LBGs with $K_s$-band detection, we find that their color ranges are
 $1<(I-K_s)<4$ and $\langle(K_s-4.5\mu m)\rangle=0.8$, lying in the
 space between LBGs and DRGs. In this respect, IRAC LBGs are close 
 to DRGs -- especially in view of stellar masses and ages, inferred
 from the optical/NIR colors. As for the stellar mass, we show that
 IRAC LBGs have average $M/L$ ratio of $\langle M/L_J \rangle \sim0.32$,
 while the average $M/L$ ratio of DRGs is $\langle M/L_K \rangle \sim0.33$.
 Since the rest-frame $J$- and $K$-band fluxes are nearly the same,
 the $M/L$ ratio of IRAC LBGs is comparable with that of DRGs. It
 assures the similarity of IRAC LBGs and DRGs.

 The fraction of IRAC LBGs in the whole LBG population is $\sim6$\%,
 and it can be drawn from the known overlap between LBGs and DRGs,
 $\sim10$\%. As is noted in Section 3.2, the fraction is a
 function of image depth -- however, we see that IRAC LBGs represent
 most of ``massive'' high-redshift galaxies which could be selected
 from Lyman break selection criteria.

 \subsection{Dust Properties of $z\sim3$ LBGs }

   \begin{figure*}
  \epsscale{1}
  \plottwo{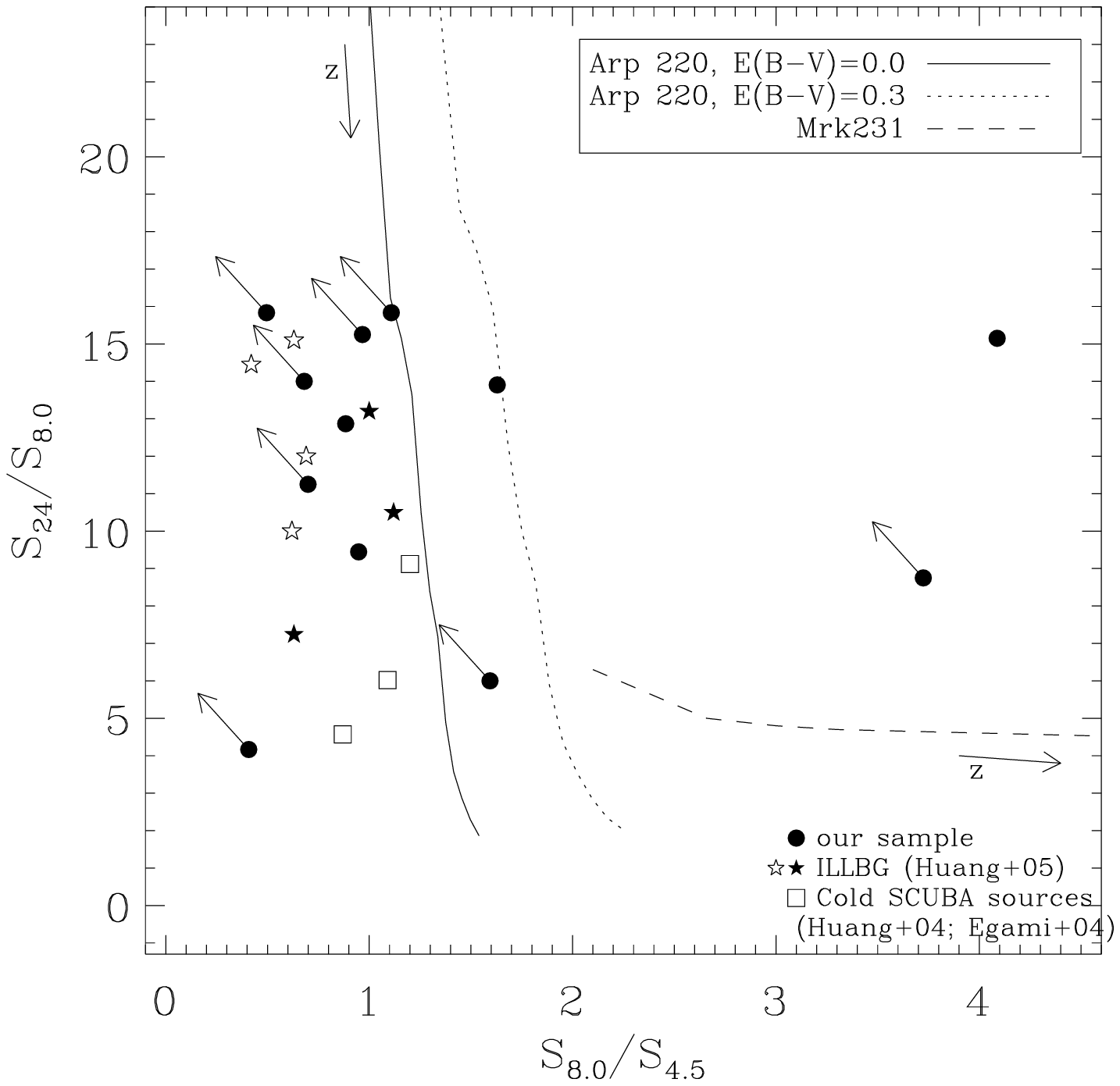}{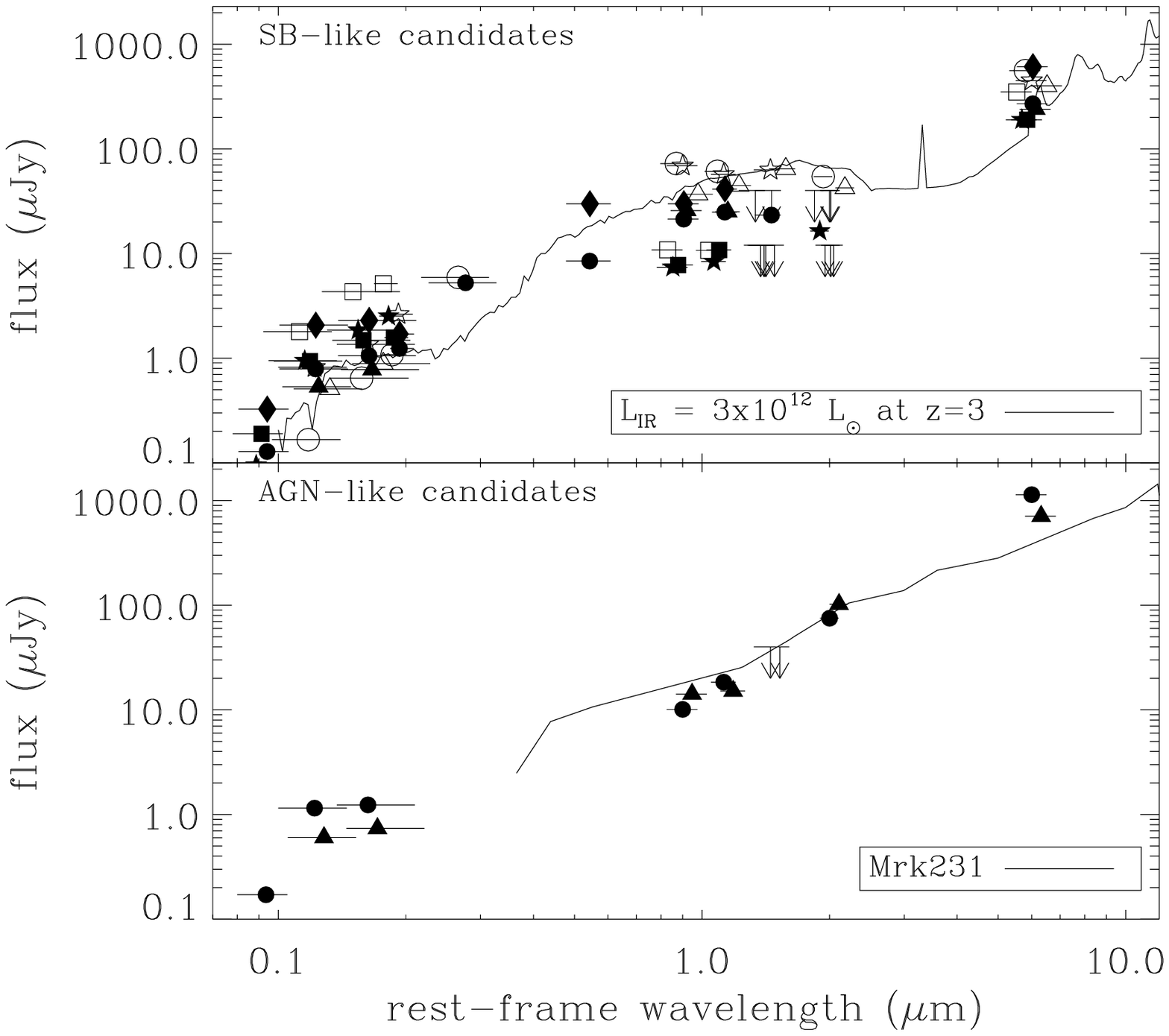}
  \caption{\label{fig:mips} \textit{Left}: The IR color-color
  diagram of IRAC LBGs detected in 24 $\mu$m.
  The lines represent tracks of three different spectral templates
  (\textit{solid} line for ULIRG Arp 220 with no additional
  reddening;\textit{dotted} line for Arp 220 with $E(B-V)=0.3$ ;
  \textit{dashed} line for dusty AGN Mrk231).
  At each line, the redshift is increasing downward and to the right.
  The 24 $\mu$m LBGs are plotted as filled circles, with arrows
  indicating the upper limit value for 8 $\mu$m flux.
  The stars indicate ILLBGs from Huang et al.(2005) that are likely
  to be starburst galaxies (filled star for spectroscopic sample and
  open star for photometric sample), and the squares represent cold
  SCUBA sources (Huang et al. 2004; Egami et al. 2004).
  Here, starburst galaxies reside in the left part of the plot and
  the AGNs are located in the bottom-right.
  \textit{Right}: The spectral energy distribution of 24 $\mu$m LBGs.
  Photometric points of each objects are converted to the rest-frame
  value using photometric redshift. They are similar to starburst galaxies
  (\textit{top}), or AGN (\textit{bottom}).
  Overplotted lines are IR galaxy template with
  $L_{IR}=3\times10^{12} L_{\odot}$(\textit{top}, Chary \& Elbaz 2001),
  empirical SED of AGN Mrk 231 (\textit{bottom}).
  }
  \end{figure*}

 \subsubsection{24 $\mu$m detected IRAC LBGs}

  In this subsection, we discuss properties of twelve 
 24 $\mu$m LBGs, which are the IRAC LBGs also detected in the 
 FLS MIPS image (Section 3.2).

  First, we investigate the origin of the dust emission, i.e., 
 whether it comes from AGN or starburst.
  The combination of MIPS 24 $\mu$m flux and other IRAC fluxes is
 known to be a useful indicator to weed out AGNs from star-forming 
 IR-bright galaxies (Egami et al. 2004; Ivison et al. 2004;
 Yan et al. 2004).
  Figure \ref{fig:mips}a shows the ratio of 8 $\mu$m flux to
 4.5 $\mu$m flux ($S_{8.0}/S_{4.5}$) versus the ratio of
 24 $\mu$m flux to the 8 $\mu$m flux ($S_{24}/S_{8.0}$;
 see Figure 3 of Ivison et al. 2004). Objects that have large
 $S_{8.0}/S_{4.5}$ and small $S_{24}/S_{8.0}$ are considered
 to be AGNs in this diagram.
  Although many objects in our sample have only upper limits
 in IRAC 4.5 $\mu$m, 5.8 $\mu$m or 8.0 $\mu$m, we examined whether
 the objects could be classified as AGN or starburst.
  The result shows that there are 2 possible AGN candidates among
 our 24 $\mu$m LBGs, while the rest can be classified as star-forming
 galaxies that have MIR colors similar to ILLBG or cold SCUBA sources
 in Huang et al. (2005).
  In a previous discussion, we mentioned that the bright end of 
 the UV luminosity function could be affected by AGNs. Note that  
 the two AGN-type 24 $\mu$m LBGs have $R$-band magnitude of
 23.7 mag, and 24.3 mag, therefore they are not directly related 
 to the objects consisting of the bright end of the UV-luminosity 
 function.

  In Figure \ref{fig:mips}b, we plot SEDs of the starburst-type
 and the AGN-type 24 $\mu$m LBGs. Solid lines are a SED of a
 star forming 
 galaxy with $L_{IR}=3\times10^{12} L_{\odot}$ from Chary \& Elbaz
 (2001) and the SED of AGN Mrk231.
  The top panel of Figure \ref{fig:mips}b confirms that 
 most of the 24 $\mu$m LBGs have SEDs similar to luminous infrared 
 galaxies with strong starburst activity.
  The rest-frame NIR part of their SEDs is nearly flat or only
 moderately increasing toward the longer wavelength, suggesting
 that the rest-frame NIR fluxes are mostly due to stellar light.
  The broad PAH emission features at the rest-frame 6--8 $\mu$m shift  
 to 24 $\mu$m at $z\sim3$, therefore we interpret 24 $\mu$m detection 
 to be due to PAH emissions.
  On the other hand, the two AGN-type 24 $\mu$m LBGs have a strong 
 power-law continuum typical of AGNs (Figure \ref{fig:mips}b, bottom).
  One of these AGN-like LBGs is detected in VLA 1.4 GHz with the flux 
 of 0.59 mJy, which suggests that this object is a radio-loud AGN.

  In order to derive total infrared luminosities of 24 $\mu$m LBGs,
 we assumeed the SED of M82 ($L_{IR}\sim3\times10^{10}L_{\odot}$;
 Telesco \& Harper 1980) which is found to resemble 
 SEDs of submm galaxies at high redshift
 (e.g., Lutz et al. 2005; Menéndez-Delmestre et al. 2007)
  The derived infrared luminosities
 are of order of a few $\times~10^{12}$ $L_{\odot}$ to
 $10^{13}$ $L_{\odot}$ (Table \ref{tab:com_sfr}),
 suggesting that the 24 $\mu$m LBGs are Ultraluminous Infrared Galaxies.

  \begin{deluxetable*}{lcccccccc}
  \tabletypesize{\scriptsize}
  \tablewidth{0pt}
  \setlength{\tabcolsep}{0.005in}
  \tablecaption{Comparison of UV-derived SFR and IR-derived SFR
  \label{tab:com_sfr}}
  \tablehead {
  \colhead{object ID} &
  \colhead{$24\mu$m flux} &
  \colhead{UV SFR} &
  \colhead{UV SFR$_c$} & \colhead{IR SFR} &
  \colhead{IR luminosity} &
  \colhead{$E(B-V)$}  & \colhead{A$_{1600}$} &
  \colhead{comment}
  \\
  \colhead{} & \colhead{(mJy)} & \colhead{($M_{\odot}/yr$)} & \colhead{($M_{\odot}/yr$)} &
  \colhead{($M_{\odot}/yr$)} & \colhead{($L_{\odot}$)} & \colhead{} & \colhead{(mag)} & \colhead{}
  \\
  \colhead{(1)} & \colhead{(2)} & \colhead{(3)} & \colhead{(4)} &
  \colhead{(5)} & \colhead{(6)} & \colhead{(7)} & \colhead{(8)} & \colhead{(9)}
  }
  \startdata
   c[320] FLS\_R\_J171734.5+593548 & $\phantom{+}0.14\pm0.05$ & 51.7 & 424.3 (1413.3) & 454.9 & 2.66$\times$10$^{12}$ & 0.36 & 3.59 & \nodata \\
   c[369] FLS\_R\_J171556.8+593833 & $\phantom{+}0.27\pm0.05$ & 23.8 & 146.2 (163.9) &  877.2 & 5.13$\times$10$^{12}$ & 0.21 & 2.10 & \nodata \\
   c[421] FLS\_R\_J171735.3+594137 & $\phantom{+}0.19\pm0.05$ & 35.7 & 107.1 (204.8) &  617.3 & 3.61$\times$10$^{12}$ & 0.19 & 1.90 & \nodata \\
   c[583] FLS\_R\_J171430.7+595213 & $\phantom{+}0.24\pm0.05$ & 16.9 & 383.4 (242.4) &  779.8 & 4.56$\times$10$^{12}$ & 0.29 & 2.89 & \nodata \\
   c[606] FLS\_R\_J171853.5+595325 & $\phantom{+}0.19\pm0.05$ & 47.6 & 469.1 (473.6) &  617.3 & 3.61$\times$10$^{12}$ & 0.25 & 2.49 & \nodata \\
   c[678] FLS\_R\_J171418.8+595722 & $\phantom{+}0.56\pm0.05$ & 15.9 & 955.4 (2723.5) & 1812.6 & 1.06$\times$10$^{13}$ & 0.56 & 5.59 & \nodata \\
  \hline
   o[035] FLS\_R\_J171412.0+591716 & $\phantom{+}0.61\pm0.05$ & 51.5 & 204.5 (269.3) & 1983.6 & 1.16$\times$10$^{13}$ & 0.18 & 1.80 & ?  \\
   o[217] FLS\_R\_J171642.9+585733 & $\phantom{+}0.35\pm0.05$ & 117.8 & 200.9 (155.2) & 1137.2 & 6.65$\times$10$^{12}$ & 0.03 & 0.30 & ?  \\
   o[275] FLS\_R\_J171816.7+584813 & $\phantom{+}0.40\pm0.06$ & 16.8 & 620.8 (317.6) & 1299.6 & 7.60$\times$10$^{12}$ & 0.32 & 3.19 &  \nodata \\
   o[306] FLS\_R\_J172045.2+585221 & $\phantom{+}1.14\pm0.06$ & 28.2 & 35.6 (70.8) & 3710.7 & 2.17$\times$10$^{13}$ & 0.10 & 1.00 & AGN?\tablenotemark{a} \\
   o[429] FLS\_R\_J172202.1+585414 & $\phantom{+}0.45\pm0.05$ & 30.9 & 418.7 (405.7) & 1462.1 & 8.55$\times$10$^{12}$ & 0.28 & 2.79 & \nodata \\
   o[504] FLS\_R\_J171618.4+602620 & $\phantom{+}0.71\pm0.06$ & 15.0 & 38.6 (23.8) & 2308.5 & 1.35$\times$10$^{13}$ & 0.05 & 0.50 & AGN? \\
  \enddata
  \tablecomments
  { (1) ID of the object (drawn from $R$-band catalogs of Fadda et al.(2004));
  (2) 24 $\mu$m flux in mJy; (3) uncorrected UV-derived SFR; (4) corrected
  UV-derived SFR using UV slope $\beta$ or $E(B-V)$ (see Section 5.3.1
  for more detail); (5) IR-derived SFR; (6) IR luminosity using 24 $\mu$m
  flux; (7) $E(B-V)$ derived during the SED fitting process. According to
  Calzetti et al.(2000), $E_s(B-V)=0.44E_g(B-V)$, while
  $A_{1600}=4.39E_{g}(B-V)$ for $E_{g}(B-V)$ derived using nebular gas
  emission lines; (8) $A_{1600}$ estimated from $E(B-V)$; (9) comments for
  the objects. Objects marked as ``AGN?'' are suspicious to be AGNs due to
  their SED shapes in MIR (see Figure \ref{fig:mips} in Section 5.3.1).
  Objects with ``?'' mark (o[035], o[217]) are objects with a large
  discrepancy between UV-derived SFR and IR-derived SFR, thus are
  suspected to be AGN-dominated.
  }
  \tablenotetext{a}{This object is detected in VLA 1.4 GHz image, with
  the flux of 0.59 mJy.}
  \end{deluxetable*}

   \begin{figure*}
  \plotone{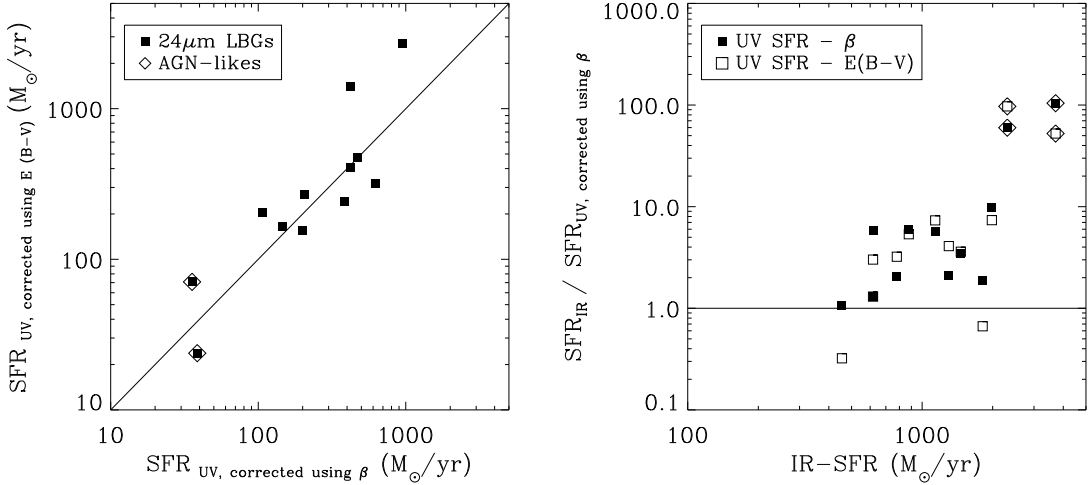}
  \caption{\label{fig:com_sfr}
  \textit{Left}: The comparison between two different extinction
  correction method to UV-derived SFR. Both method, using UV-slope
  $\beta$ or $E(B-V)$ are consistent within a factor of $\sim2$.
  \textit{Right}: The comparison between IR-derived SFR and
  extinction-corrected SFR. When IR-SFR is below $\sim800 M_{\odot}/yr$,
  extinction-corrected UV-SFR is quite consistent with IR-SFR.
  }
  \end{figure*}

 Alternatively, we tried to estimate $L_{IR}$ of 24 $\mu$m LBGs using
 infrared SED templates of Chary \& Elbaz (2001), in which 
 the MIR flux correlates with $L_{IR}$ as $L_{IR} \propto 
 L_{6.7\mu m}^{1.62}$.
 With such a method, we obtain
 unreasonably high values of $L_{IR}\sim10^{14} - 10^{15}~L_{\odot}$
 for 24$\mu$m LBGs.
 If 24 $\mu$m LBGs were really as bright as $L_{IR} \simeq 10^{14} - 
 10^{15}~L_{\odot}$, their 850 $\mu$m flux would be well above
 10 mJy, and they shoud be detectable with submm observations. 
 However, the existing submm data with a partial coverage
 of FLS show no detection of 24 $\mu$m LBGs providing
 the flux limit of $< 5$ mJy
 (Sawicki \& Webb 2005; Kim et al. 2007, in preparation).
 Therefore, we conclude that cautions are needed when deriving 
 $L_{IR}$ of $z \sim 3$ objects using the correlation between
 MIR flux and IR flux such as those found in Chary \& Elbaz 
 (2001). 
  
  We now investigate the dust properties of 24 $\mu$m LBGs by 
 comparing SFRs derived from UV luminosity (corrected for
 dust extinction) and also from IR luminosity.
  When deriving SFRs from UV luminosity, two methods are used to 
 correct for the dust extinction.
  One method uses the UV slope $\beta$ defined as 
 $f_{\lambda} \propto {\lambda}^{\beta}$ at 1500--2800$\mbox{\AA}$ 
 (Meurer, Heckman, \& Calzetti 1999). Since there are no spectroscopic
 data for our LBG sample, we measured $\beta$ from the model
 spectral template which is found to be the best-fit model using
 optical ($u,g,R$ and $i$) photometry only.
  The other method uses the $E(B-V)$ value derived from the
 SED-fitting and a known correlation between the UV extinction
 and $E(B-V)$ ($A_{1600}=4.39E_g(B-V)$; Calzetti et al. 2000).
  Here, our $E(B-V)$ derived from stellar continuum is related with 
 nebular line-derived $E_g(B-V)$ as $E_s(B-V)=0.44E_g(B-V)$ 
 (When we use $E(B-V)$ value, we mean $E_s(B-V)$ hereafter).  
  The conversion formula from luminosity to star formation rates 
 are taken from Kennicutt (1998):

\begin{center}
  \begin{tabular}{l}
  $ SFR_{IR} (M_{\odot}/yr)=1.71\times10^{-10} L_{IR}(L_{\odot}) $ \\
  $ SFR_{UV} (M_{\odot}/yr)=1.4\times10^{-28} L_{UV} (ergs/s/Hz)  $
  \end{tabular}
\end{center}

  The result is presented in Table \ref{tab:com_sfr} and Figure
 \ref{fig:com_sfr}. As is shown in the Table \ref{tab:com_sfr}, 
 the UV-derived SFRs from two methods are consistent with each other,
 within a factor of 2 in most cases. 
  Between the extinction-corrected UV-derived SFRs and IR-derived SFR,
 Figure \ref{fig:com_sfr} shows that these quantities are consistent
 with each other within a factor of a few, and with the median ratios
 of $\sim2$ when $SFR_{IR} \lesssim 800 M_{\odot}/yr$ 
 or $L_{IR} < 5\times10^{12} L_{\odot}$.
  Note that we ignore two AGN-type objects in this comparison. 
  However, for galaxies with $L_{IR} > 5\times10^{12}L_{\odot}$ or
 $SFR_{IR} \gtrsim 800 M_{\odot}/yr$, we find that 
 extinction-corrected $SFR_{UV}$s from the both $\beta$-correction
 and the $E(B-V)$ methods are either systematically  
 lower than $SFR_{IR}$ by a factor of a few or have large 
 scatter with respect to $SFR_{IR}$. 
  If the $SFR_{IR}$ values represent the true star formation rates, 
 this result suggests that the extinction correction at UV 
 tend to be underestimated when $L_{IR} > 5 \times 10^{12} L_{\odot}$,
 consistent with the previous findings where it has been found that
 the UV-slope method appears to systematically underestimate the extinction
 correction in UV for the most IR luminous galaxies 
 (Papovich et al. 2006; Reddy et al. 2006).

  Table \ref{tab:com_sfr} shows that the dust-free SFR of LBGs
 using 24 $\mu$m flux is of order of 
 a few hundred to a few thousand $M_{\odot}/yr$.
 Assuming that IR luminosity represents the total star forming activity, 
 we derive the extinction correction at 1600$\mbox{\AA}$ to be
 $A_{1600} = 2$--4.5 mag for 24 $\mu$m LBGs, by comparing the
 $SFR_{IR}$ and $SFR_{UV,uncorrected}$.
  Previous works have found that the LBGs at $z\sim3$ have a
 median extinction value of $\langle A_{1600}\rangle=1.0$ mag, 
 distributed over $A_{1600}=0$--4 mag (Adelberger \& Steidel 2000).
  Comparing the extinction values of 24 $\mu$m LBGs with the above 
 values, we conclude that 24 $\mu$m LBGs are the dustiest population 
 among LBGs.

  We also derived $A_{1600}$ from $E(B-V)$. Since we use $E(B-V)$ 
 to correct for the dust extinction of UV luminosity of LBGs 
 in the following analysis, we mention here how reliable 
 the extinction correction using $E(B-V)$ would be.
  Mostly, there are $\sim2$ mag scatter in $A_{1600}$. First, the 
 comparison between $SFR_{IR}$ and $SFR_{UV}$ suggest that 
 the extinction correction using $E(B-V)$ is reasonably accurate 
 to a factor of $\sim$ 2.
  Additional uncertainty exists in the relation between 
 $A_{1600}$ and $E(B-V)$ observationally (Calzetti et al. 2000)
 and theoretically (Witt \& Gordon 2000). Calzetti et al.(2000) mention
 that the linear relation between $E(B-V)$ and $A_{1600}$ represents 
 the lowest envelope of the relation, suggesting the derived 
 $A_{1600}$ might underestimate the dust extinction by a factor of a few.

  \subsubsection{Dust property vs. stellar mass}

\begin{figure}
  \epsscale{1.2}
  \plotone{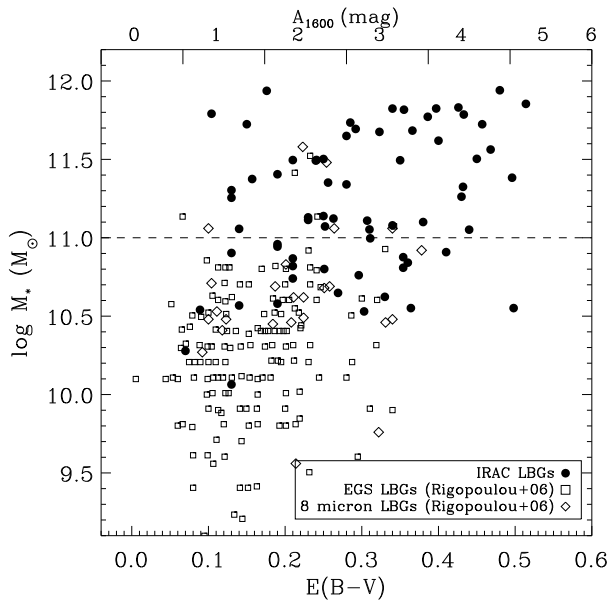}
  \caption{\label{fig:ebv}
  The $E(B-V)$ versus the stellar mass.
  Also, overplotted on the figure are the data points from
  Rigopoulou et al.(2006). The median value of dust extinction in IRAC LBGs
  is $\langle E(B-V)\rangle=0.29$. On average, there is a correlation
  between dust extinction and stellar mass with a correlation
  coefficient $r=0.51$.}
  \end{figure}

  In Figure \ref{fig:ebv}, we show a plot of dust extinction $E(B-V)$
 versus stellar mass. When our massive ($>10^{11} M_{\odot}$) LBGs are
 plotted together with less massive LBGs from Rigopoulou et al.(2006), 
 there is a weak, but positive correlation between $log M_*$ and
 $E(B-V)$ with the correlation coefficient of $r=0.51$. Thus, 
 Figure \ref{fig:ebv} suggests that dustier LBGs tend to be more massive.
 It has been argued that LBGs with intrinsically higher bolometric
 luminosity (UV+IR) are more massive (Reddy et al.
 2006; $1.5<z<2.6$ optically selected LBGs). Large amount of dust
 in LBG implies high infrared luminosity that leads to high bolometric
 luminosity, therefore our result is consistent with 
 previous results.

  The median value for the dust extinction in IRAC LBGs is 
 $\langle E(B-V) \rangle \simeq 0.29$. This is larger than 
 $\langle E(B-V) \rangle \simeq 0.15$--0.20 of all LBG population
 (Adelberger \& Steidel, 2000) and similar to ILLBGs of 
 $\langle E(B-V) \rangle = 0.354$ (Rigopoulou et al. 2006).
 We already mentioned in Section 5.3.1 that the mean value for dust
 extinction is $E(B-V)=0.28$ for 24$\mu$m LBGs.
 IRAC LBGs and 24$\mu$m LBGs share the same range
 of dust-extinction, while the infrared luminosities of 24$\mu$m
 LBGs are higher than those of IRAC LBGs. By stacking the 
 24 $\mu$m images of LBGs with no individual detection in 
 24 $\mu$m image, we see that the average IR luminosities of 
 IRAC LBGs are slightly lower, but close to those of 24$\mu$m LBGs 
 within a factor of a few (presented in Section 5.3.3 in more detail).

  \subsubsection{Average infrared luminosity of LBGs}

  24 $\mu$m-detected LBGs are probably the most infrared-luminous
 galaxies among all LBGs, with the total infrared luminosity of 
 $\gtrsim$ a few $\times~ 10^{12} L_{\odot}$. Since the majority of LBGs are
 not detected in 24 $\mu$m, we examined their IR properties by stacking
 MIPS 24 $\mu$m images.

\begin{figure}
 \plotone{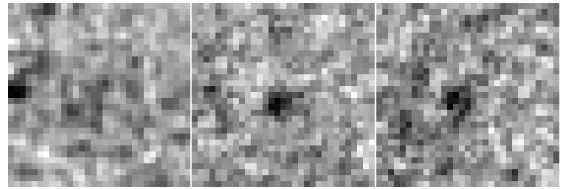}
 \caption{
 \label{fig:posimg} \textit{Left} is an example of 24 $\mu$m postage
 image of IRAC LBG that is not detected in MIPS 24 $\mu$m image.
  We also present the median-stacked images of 20 IRAC LBGs without
 24 $\mu$m detection in the verification strip (\textit{middle}),
 and 49 IRAC LBGs without 24 $\mu$m detection in the main field
 (\textit{right}). The size of each image is $40\arcsec\times40\arcsec$.
 The stacked image shows a clear signal over 8 $\sigma$.
 }
 \end{figure}

  First, we examined IR properties of IRAC LBGs.
 We stacked 24 $\mu$m images of 20 IRAC LBGs in the MIPS FLS
 verification strip and 49 IRAC LBGs in the main field,
 both without 24 $\mu$m detection (Figure \ref{fig:posimg}).
 The stacked images show a clear existence of emission in 24 $\mu$m.
 When measured upon the stacked image, the signal to noise ratios
 are $\sim8$ in both the verification and the main fields.
 The average fluxes of these LBGs, therefore,
 are 60.4 $\mu$Jy for 24 $\mu$m-undetected IRAC LBGs in the 
 verification strip, and 70.3 $\mu$Jy for 24 $\mu$m-undetected 
 IRAC LBGs in the main field.
  Converting the expected 24 $\mu$m flux to the total infrared
 luminosity assuming M82 SED, we find that the expected
 IR luminosity of 24 $\mu$m undetected IRAC LBGs is about
 $\sim1.1\times10^{12} L_{\odot}$ in the verification strip, and
 $\sim1.3\times10^{12} L_{\odot}$ in the main field.
  The corresponding star formation rate is
 180--220$M_{\odot}/yr$. The estimated 24 $\mu$m flux of 
 60.4--70.3$\mu$Jy for IRAC LBGs is comparable with the 60 $\mu$Jy
 cut for ILLBGs defined in Huang et al.(2005), who suggest
 the ILLBGs occupy 5\% of the total LBGs. The fraction of our
 IRAC LBGs (63) in the whole LBGs (1088) is similar with their
 value.

  With the same stacking method, we also estimated the average 
 infrared flux of the LBGs not detected in IRAC. The 24 $\mu$m
 images of 300 LBGs in the MIPS verification field with no IRAC
 detection were stacked, showing a marginal detection of 
 $S/N\sim3$ or the average flux of $6~\mu$Jy.
  The estimated infrared luminosity of these LBGs is of order of
 $1.1\times10^{11} L_{\odot}$, or the star formation rate of
 $19 M_{\odot}/yr$ according to the conversion relation.
 This value is consistent with the SFR, $10 - 100 M_{\odot}/yr$,
 of typical LBGs as mentioned in the Introduction.
 At this relatively low infrared luminosity range, the result from 
 Chary \& Elbaz template fitting is not much different from the M82
 scaling method; the result is
 $\langle L_{IR} \rangle = 1.4\times10^{11} L_{\odot}$, the star
 formation rate of $24 M_{\odot}/yr$.

 \subsection{Implication for Galaxy Formation}

 \subsubsection{Number density of massive LBGs}

\begin{figure}
  \epsscale{1.2}
 \plotone{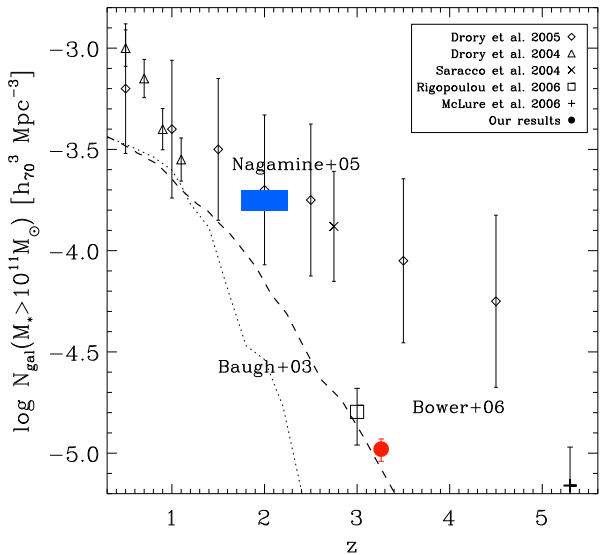}
 \caption{\label{fig:down_numd} The number density of galaxies with
  stellar masses $>10^{11}M_{\odot}$ as a function of redshift. The
  red filled circle shows our result for massive LBGs, which is quite
  comparable with the result of massive LBGs at the similar redshift
  range by Rigopoulou et al.(2006). Data points from other observations
  are plotted with different symbols, and the prediction from
  recent semi-analytic and hydrodynamic models of galaxy formation are
  overplotted (\textit{dotted} line from Baugh et al. 2003;
  \textit{dashed} line for Bower et al. 2006; blue filled rectangle
  from Nagamine et al. 2005).
 }
 \end{figure}

  The number density of massive galaxies at high redshift can show 
 the straightforward evidence of early formation of massive systems.
  In Figure \ref{fig:down_numd}, we plot the number density of
 massive ($M > 10^{11} M_{\odot}$) LBGs in filled circle, and compare
 the value with other observational results (Drory et al. 2005;
 Saracco et al. 2004; Rigopoulou et al. 2006; McLure et al. 2006)
 and the predictions from several hierarchical galaxy formation models
 such as semi-analytic models of Baugh et al.(2003; dotted line), 
 Bower et al.(2006; dashed line), and the hydrodynamic simulation
 of Nagamine et al.(2005; large solid rectangle).
  Note that the more recent models (Bower et al. 2006) overcome
 shortcomings of the earlier model of underestimating the number
 density of massive galaxies at high redshift (e.g., Baugh et al. 2003),
 and are successfully predicting the observational constraints at
 $z \sim 2$.

  We find that the number density of our LBGs with mass greater than
 $10^{11} M_{\odot}$ is $\Phi=(1.05\pm0.15)\times10^{-5} Mpc^{-3}$.
  Our value is consistent with the value presented in Rigopoulou et al.
 (2006), who have performed a similar study for LBGs over a smaller area.
 If we adopt $\sim20$\% as the fraction of LBGs among massive galaxies
 (van Dokkum et al. 2006), the estimated number density of all massive
 galaxies would increase to $\Phi=(5.25\pm0.75)\times10^{-5} Mpc^{-3}$,
 consistent with those from NIR-selected sample of Drory et al.(2005).
 QSO contamination at optically bright LBGs is not likely to affect the
 number density of massive LBGs since they comprise a small fraction of
 massive LBGs.

  Comparison of our result with the model predictions in Figure
 \ref{fig:down_numd} shows that the most up-to-date semi-analytic model
 (e.g., Bower et al. 2006) still underpredicts the number density of
 massive galaxies, although the discrepancy between the model and the
 observational constraints is now much reduced. Our result for the
 number density has uncertainty of a factor of a few, due to the Poisson
 errors and the small fraction that UV-selected galaxies comprise for 
 the entire massive population. However, considering that the comparison
 is made for objects at the very massive end of the mass function, the
 discrepancy could be narrowed without too much difficulty with a slight
 tweaking of model parameters, or by improving observational constraints 
 with a better number statistics and measurements.

 \subsubsection{Contribution of massive LBGs to the total SFR density}

  In the downsizing scenario of galaxy formation, the 
 star formation activity occurs early in massive galaxies, and late
 in less massive galaxies. Therefore, such a model predicts that the
 relative contribution of the star formation activity from galaxies
 with different masses should evolve as a function of redshift,
 with the star formation occurring more in massive galaxies at the
 higher redshift.

  In order to examine the contribution of massive LBGs to the total star
 formation rate at $z\sim3$, we calculated the instantaneous star
 formation rate using the UV luminosity function presented in 
 Section 5.1, and correcting it for the dust extinction.
  The FUV luminosity density is derived by 
 integrating the luminosity function over the magnitude interval
 of the survey,  
  \[ L_{FUV}=\int_{M_{min}}^{M_{max}} L(M) \Phi(M) dM  \]
 while $M_{min}$ and $M_{max}$ indicate the minimum and maximum absolute
 FUV magnitudes.

  The estimated luminosity density is then converted to the star
 formation rate density using the conversion formula from 
 Kennicutt (1998) assuming the Salpeter IMF.
 For the whole LBGs, for which we have
 an observational constraint on the faint-end slope of 
 UV luminosity function from studies extending to
 the fainter limits, we get the star formation rate
 density value (before dust extinction correction)
 of $\rho_{*} = 4.0 \times 10^{-2} M_{\odot}\,yr/Mpc^{3}$
 by using $M_{min}=-\infty$ and $M_{max}=-10$.
  For the star formation rate density of massive 
 ($>10^{11} M_{\odot}$) LBGs, we get a conservative
 estimate of $\rho_{*}=2.3\times10^{-4} M_{\odot}/yr/Mpc^3$
 by adopting $M_{min}=-23.5$ and $M_{max}=-21.0$ of our 
 survey limit, since the faint-end slope of UV luminosity
 function of massive LBGs is not well constrained. 

  Correction for the dust extinction is done as the followings.
  According to the discussion in Section 5.3.1, 
 the extinction parameter $E(B-V)$ is a reasonably accurate 
 measure of the dust extinction, except for IR-bright objects with 
 very high star formation rate. For the SFR 
 density from whole LBGs, we adopt the average dust extinction of 
 $\langle E(B-V)\rangle=0.15$ (Adelberger \& Steidel 2000). This
 average $E(B-V)$ value increases the derived SFR density by a factor 
 of $\sim3.94$ (Calzetti et al. 2000; $A(1600 \mbox{\AA}) = 4.39 E(B-V)$),
 therefore the extinction-corrected SFR density of the whole 
 LBGs at $z\sim3$ is $\rho_{*}=1.6\times10^{-1} M_{\odot}/yr/Mpc^3$.
  For the massive LBGs, we use the median $E(B-V)$ value of massive
 LBGs, $\langle E(B-V) \rangle=0.29$ (Figure \ref{fig:ebv}),
 or an extinction correction factor of 14.3.
 Therefore, the corrected SFR density of massive LBGs is 
 $\rho_{*}=3.3\times10^{-3} M_{\odot}/yr/Mpc^3$.
 As discussed in section 5.3.1,
 the uncertainty in the extinction correction factor 
 for 24 $\mu$m LBGs is about a factor of 2 to 3, but could be 
 larger than that for ULIRG-type objects. We adopt the error bar 
 of a factor of a few as a rough estimate of the SFRD uncertainty.

  The star formation rate density value of
 $\rho_{*}=3.3\times10^{-3} M_{\odot}/yr/Mpc^3$ 
 can be considered as a lower limit, since the above calculation
 do not include the contribution from massive LBGs with
 $M_{UV} > -21$. The inclusion of the population fainter than our
 survey limit into the calculation requires an assumption on the
 faint-end slope of the UV luminosity function. If we adopt a fiducial
 value of $\alpha=-1.6$ as the faint end slope, we get the star
 formation rate density of $1.6 \times 10^{-2} M_{\odot}/yr/Mpc^{3}$
 by adopting $M_{max} = -10$.  This is likely to be an upper limit
 of the star formation rate of massive galaxies, uncertain at a factor
 of roughly a few.  Also, note that the derived SFR would decrease by
 a factor of $\sim0.25$ dex if we adopt the Chabrier IMF (Chabrier 2003)
 instead of the Salpeter IMF (Dahlen et al. 2007).

  \begin{figure}
  \epsscale{1.2}
 \plotone{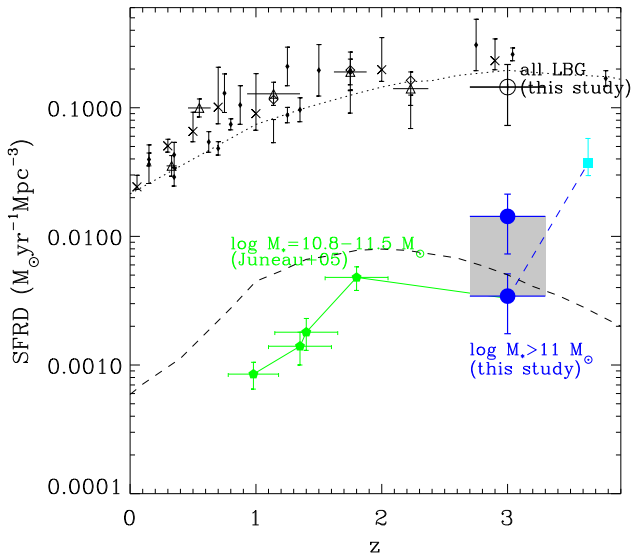}
 \caption{\label{fig:down_sfrd}
  { The star formation rate density derived from LBGs
 in FLS. The open/filled circle indicates the star formation
 rate density from whole/massive ($> 10^{11} M_{\odot}$) LBGs. The
 points are corrected for dust extinction, using average $E(B-V)$
 value ($\sim4$ and $\sim14$ for whole/massive LBGs).
 Considering the contribution from LBGs below $R$-band magnitude
 limit of our study, the estimated star formation rate density from
 massive LBGs resides in the shaded box.
  Green filled pentagons are the SFR density from galaxies whose
 stellar masses are $10^{10.8} M_{\odot} < M_* < 10^{11.5} M_{\odot}$,
 drawn from Juneau et al.(2005). Other symbols indicate the total SFR
 density at the corresponding redshift (\textit{open triangle/diamond}:
 Dahlen et al. 2007; \textit{crosses}: Schiminovich et al. 2006;
 \textit{filled diamond}: Hopkins 2004).
 Cyan filled square is the estimate of the SFR density at $z=3.6$
 using nearby massive SDSS galaxies (Panter et al. 2006).
 Overplotted \textit{dotted}/\textit{dashed} lines are the prediction
 of the evolution of star formation rate density from all/massive
 galaxies along redshift (Bower et al. 2006).
 } }
 \end{figure}

  Figure \ref{fig:down_sfrd} shows the SFR density of the LBGs 
 compared to those of galaxies at different redshifts 
 from other works.
  Various SFR densities in the figure are based on the 
 UV luminosity, although sample selection methods are different.
  What is interesting in the figure is the contribution of massive 
 galaxies to the total star formation rate density from $z<2$ 
 (Juneau et al. 2005) and $z\sim3$ (our result).
  A direct comparison of our result versus Juneau et al. (2005)
 is possible, since galaxies contributing to the most SFR in
 the $K$-band selected sample of Juneau et al.(2005) are star-forming
 galaxies which also appear to be UV-bright (cf. Burgarella et al.
 2006). Despite the slight difference in stellar mass range
 ($10^{10.8}M_{\odot} < M_* < 10^{11.5}M_{\odot}$ in 
 Juneau et al. 2005; $M_* > 10^{11} M_{\odot}$ in our study),
 our result is consistent with the view that the contribution of
 massive galaxies to the total SFR has steadily decreased from
 $z\sim3$ to the present, and possibly providing toward an even
 higher contribution at $z \sim 3.6$ as suggested by Panter et al.
 (2006). On the other hand, a large uncertainty in our data point 
 does not exclude a possible peak SFR activity for massive galaxies
 at $z \sim 2$ as predicted in a semi-analytical model of
 Bower et al. (2006; dashed line). Clearly, a better constraint
 is needed by extending the study to the fainter limit, and 
 placing more constraints on the extinction within high-redshift
 galaxies. 

  Our SFR density at $z\sim3$ does not include the contribution 
 from the UV-faint population that cannot be selected using 
 Lyman break technique, such as DRGs or submm galaxies.
  The contribution of massive galaxies to the star formation at
 $z = 3$ can go up even more if we include the contribution 
 from the galaxies that are heavily extinguished by dust.
  The number density of submm galaxies is very small, 
 $8.9\times10^{-6} Mpc^{-3}$ at $1.8<z<3.6$ (Tecza et al. 2004).
  Although there might be an overlap of order of 50\% between 
 LBGs and submm galaxies (e.g., Chapman et al. 2005), 
  the submm galaxies can have star formation rates as large as
 1000 $M_{\odot}/yr$, therefore their SFR density can be almost 
 comparable to the SFR density of massive LBGs.
  The contribution of DRGs is rather difficult to estimate. 
 F\"orster-Schreiber et al.(2004) suggested that the median
 value of the star formation rate for DRGs in HDF-S is 
 120 $M_{\odot}/yr$, and their number density is about an order of 
 magnitude greater than submm galaxies (van Dokkum et al. 2006).
  From the above fact, one might argue that the SFR density of DRG 
 is comparable to that of massive LBGs, but such an argument 
 is not valid since these two populations overlap with 
 each other as mentioned in Section 5.2.2. 

  In any case, our result suggests that the star formation activity
 in massive galaxies is a dominant process at $z = 3$ compared
 to the universe at $z < 1$, supporting the downsizing picture
 of the galaxy formation.

\section{CONCLUSION}

 We have selected and studied LBGs at $z\sim3$ 
 in the \textit{Spitzer} First Look Survey area, using the
 multi-wavelength datasets consisting of deep $u, g, R,$ and
 $i^\prime$, $J$, $K_s$-band and the \textit{Spitzer} IRAC
 and MIPS images. In total, we have found 1088 LBGs with
 $R=22.0$--24.5 mag over the $\sim2.63$ deg$^2$ area.
 The wide area coverage enables us to select a large number
 of bright LBGs with MIR fluxes which are important for
 studying various properties of rare, massive LBGs.

 Particularly, we concentrated on the properties of LBGs detected
 in IRAC 3.6$\mu$m over 3$\sigma$ (6$\mu$Jy in verification 
 strip, 12$\mu$Jy in main field). These ``IRAC LBGs'' are on 
 average massive/old/infrared sub-population of whole LBGs.
 Nearly $\sim70$\% of IRAC LBGs are more massive than $10^{11} M_{\odot}$.
 IRAC LBGs with the largest stellar mass have the reddest
 (optical-MIR) color, which is indicative of old stellar population.
 $M/L$ ratio of the galaxies dominated by old stellar population
 is constant with little scatter. On the other hand, IRAC LBGs
 with large ongoing star formation increase the scatter in $M/L$.

 Among these IRAC LBGs, 12 LBGs were detected in 24 $\mu$m image.
 The infrared luminosity of the LBGs with individual 24 $\mu$m
 detection suggests a high star formation rate of $\sim1000M_{\odot}/yr$
 occurring in these systems. The $E(B-V)$ of LBGs, indicating the amount
 of dust extinction within the uncertainty of a factor of a few,
 has weak but existing correlation with the stellar mass. Dustier
 LBGs are more massive, and this again ensures we see the
 massive/infrared end of LBG population in the IRAC LBGs. 

 With the photometric redshifts of the LBGs,
 the rest-frame UV luminosity function is constructed.
 The derived luminosity function is consistent with previous studies,
 but with a much improved number statistics. We also construct
 the UV luminosity function of massive LBGs ($>10^{11} M_{\odot}$),
 from which we estimate the SFR density in massive systems at
 high redshift. The star formation rate density from all LBGs 
 at $z\sim3$, calculated in the survey magnitude interval, is
 $\rho_{*}=1.6\times10^{-1} M_{\odot}/yr/Mpc^3$,
 while the star formation rate density from massive LBGs is
 $\rho_{*}=3.3\times10^{-3}\sim1.6\times10^{-2} M_{\odot}/yr/Mpc^3$.
 The contribution of the massive systems to the global star formation
 at $z\sim3$ is significantly large compared to the
 case of lower redshifts. This finding suggests that the shift of
 star formation activity from massive systems to the smaller systems
 as the universe ages, interpreted as known ``downsizing'' 
 of the galaxy formation and evolution.

\acknowledgements
  We thank the FLS members for their support on this program.
  This work was supported by the grant No. R01-2005-000-10610-0 from
 the Basic Research Program of the Korea Science \& Engineering 
 Foundation, and the university-institute
 cooperative research fund from the Korea Astronomy and Space
 Science Institute.
  We also acknowledge the support from
 the Frontier Physics Research Division of the 
 Brain Korea 21 program at Seoul National University.
  We thank an anonymous referee for many useful comments.

\clearpage

\end{document}